\documentclass[12pt]{article}
\usepackage{eucal}
\let\mysub=\subset

\DeclareFontFamily{T1}{calligra}{}
\DeclareFontShape{T1}{calligra}{m}{n}{<->s*[1.44]callig15}{}
\DeclareMathAlphabet\mathcalligra   {T1}{calligra} {m} {n}
\DeclareMathAlphabet\mathzapf       {T1}{pzc} {mb} {it}
\DeclareMathAlphabet\mathchorus     {T1}{qzc} {m} {n}
\DeclareMathAlphabet\mathrsfso      {U}{rsfso}{m}{n}
\DeclareMathAlphabet\mathfrcal      {T1}{frcursive}{m}{it}
\DeclareFontFamily{T1}{frcursive}{}
\DeclareFontShape{T1}{frcursive}{m}{n}{<->s*[1.44]callig15}{}
\DeclareMathAlphabet\mathfrcal      {T1}{frcursive}{m}{it}

\usepackage{amsmath}
\usepackage{amssymb}
\usepackage{graphicx}
\usepackage{xcolor}
\numberwithin{equation}{section}
\usepackage{array}
\usepackage{mathtools}
\usepackage{dsfont}
\usepackage{mathrsfs}
\usepackage{tikz}\usetikzlibrary{matrix,fit}
\usepackage{varwidth}
\usepackage{enumerate}

\usepackage[margin=1in]{geometry}

\usepackage{mathabx}
\usepackage{empheq}

\usepackage{hyperref}
\usepackage{setspace}
\usepackage[square,sort,comma,numbers]{natbib}

\usepackage{slashed}
\usepackage{upgreek}
\usepackage{appendix}

\usepackage{tabstackengine}
\fixTABwidth{T}

\newdimen\mytextwidth
\newcommand\rem[2][cyan!40!green]{\noindent\nobreak\hfil\penalty1000\hfilneg
\mytextwidth=\linewidth\advance\mytextwidth by 2mm%
\begin{tikzpicture}[baseline=-\the\dimexpr\fontdimen22\textfont2\relax]\node[outer sep=0pt,draw=black,fill=#1,fill opacity=1,text opacity=1,rectangle,rounded corners]{\begin{varwidth}{\mytextwidth}\textcolor{white}{#2}\end{varwidth}};
\end{tikzpicture}\allowbreak%
}

\newcommand\whiterem[2][white!]{\noindent\nobreak\hfil\penalty1000\hfilneg
\mytextwidth=\linewidth\advance\mytextwidth by 2mm%
\begin{tikzpicture}[baseline=-\the\dimexpr\fontdimen22\textfont2\relax]\node[outer sep=0pt,draw=black,fill=#1,fill opacity=1,text opacity=1,rectangle,rounded corners,line width=1.5pt]{\begin{varwidth}{\mytextwidth}\textcolor{black}{#2}\end{varwidth}};
\end{tikzpicture}\allowbreak%
}

\newcommand{\dd}{\partial}
\newcommand{\bd}{\overline{\partial}}
\newcommand{\CP}{\mathds{CP}}
\newcommand{\CC}{\mathds{C}}

\renewcommand{\bar}{\overline}
\renewcommand{\tilde}{\widetilde}

\newcommand{\Rc}{\mathcal{R}}

\newcommand{\bea}{\begin{equation}}
\newcommand{\eea}{\end{equation}}
\newcommand{\bear}{\begin{eqnarray}}
\newcommand{\eear}{\end{eqnarray}}
\newcommand{\bearr}{\begin{eqnarray*}}
\newcommand{\eearr}{\end{eqnarray*}}

\newcommand{\appendixnumberline}[1]{Appendix #1.\space}

\let\oldappendix\appendix
\makeatletter
\renewcommand{\appendix}{%
  \addtocontents{toc}{\let\protect\numberline\protect\appendixnumberline}%
  \renewcommand{\@seccntformat}[1]{\large\bfseries Appendix . }%
  \oldappendix
}
\makeatother

\usepackage{mdframed}

\newmdenv[
  topline=false,
  bottomline=false,
  rightline=false,
  linewidth=2pt,
  skipabove=\topsep,
  skipbelow=\topsep
]{siderules}

\newmdenv[
  topline=false,
  bottomline=false,
  linewidth=2pt,
  skipabove=\topsep,
  skipbelow=\topsep
]{siderulesright}

\makeatletter
\renewcommand{\@seccntformat}[1]{\csname the#1\endcsname.\quad}
\makeatother

\newcommand\scalemath[2]{\scalebox{#1}{\mbox{\ensuremath{\displaystyle #2}}}}

\usepackage{setspace}
\usepackage{xpatch}

\makeatletter
\renewcommand{\@chap@pppage}{%
  \clear@ppage
  \thispagestyle{plain}%
  \if@twocolumn\onecolumn\@tempswatrue\else\@tempswafalse\fi
  \null\vfil
  \markboth{}{}%
  {\centering
   \interlinepenalty \@M
   \normalfont
   \MakeUppercase \appendixpagename\par}%
  \if@dotoc@pp
    \addappheadtotoc
  \fi
  \vfil\newpage
  \if@twoside
    \if@openright
      \null
      \thispagestyle{empty}%
      \newpage
    \fi
  \fi
  \if@tempswa
    \twocolumn
  \fi
}
\makeatother

\usepackage{tocloft}

\let \savenumberline \numberline
\def \numberline#1{\savenumberline{#1.}}

\usepackage{etoolbox}
\patchcmd{\tableofcontents}{\@starttoc}{\vspace{-0.3cm}\@starttoc}{}{}

\usepackage{hyperref}

\usepackage{titlesec}

\titleformat*{\section}{\large\bfseries}
\titleformat*{\subsection}{\normalsize\bfseries}
\titleformat*{\subsubsection}{\normalsize\bfseries}
\titleformat*{\paragraph}{\large\bfseries}
\titleformat*{\subparagraph}{\large\bfseries}
\titlespacing{\author}{-5pt}{-5pt}{-5pt}[-5pt]

\makeatletter
\renewcommand\subsubsection{\@startsection{subsubsection}{3}{\z@}%
                                     {-3.25ex\@plus -1ex \@minus -.2ex}%
                                     {-1.5ex \@plus -.2ex}
                                     {\normalfont\normalsize\bfseries}}
\renewcommand\subsection{\@startsection{subsection}{3}{\z@}%
                                     {-3.25ex\@plus -1ex \@minus -.2ex}%
                                     {-1.5ex \@plus -.2ex}
                                     {\normalfont\normalsize\bfseries}}                                     
\makeatother

\begin{document}

\title{\vspace{-1.0cm} Quantum flag manifold $\sigma$-models\\ \hspace{-0.3cm}{\Large \emph{and}} Hermitian Ricci flow}
\author{Dmitri Bykov$^{1,2, 3}$\footnote{Emails:
bykov@mpp.mpg.de, bykov@mi-ras.ru, dmitri.v.bykov@gmail.com}
\\  \vspace{-0.3cm}  \\
{\small $^1$ Max-Planck-Institut f\"ur Physik, F\"ohringer Ring 6, D-80805 Munich, Germany} \vspace{-0.2cm}\\ \vspace{-0.2cm}
{\small $^2$ Arnold Sommerfeld Center for Theoretical Physics,}\\ {\small Theresienstrasse 37, D-80333 Munich, Germany} \vspace{-0.2cm} \\ \vspace{-0.2cm}
{\small $^3$ Steklov
Mathematical Institute of Russ. Acad. Sci.,}\\ {\small Gubkina str. 8, 119991 Moscow, Russia \;}
}

\date{}

\begin{flushright}    
  {\small
    MPP-2020-103 \\
    LMU-ASC 30/20
  }
\end{flushright}

{\let\newpage\relax\maketitle}

\maketitle

\vspace{-0.5cm}
\begin{siderulesright}
We show that flag manifold $\sigma$-models (including $\CP^{n-1}$, Grassmannian models as special cases) and their deformed versions may be cast in the form of gauged bosonic Thirring/Gross-Neveu-type systems. Quantum mechanically the gauging is violated by chiral anomalies, which may be cancelled by adding fermions. We conjecture that such models are integrable and check on some examples that the trigonometrically deformed geometries satisfy the generalized Ricci flow equations.
\end{siderulesright}

\tableofcontents

\section{Introduction and main results}

\subsection[Previous work]{Previous work.}
The present paper is part of a series, where we study two-dimensional $\sigma$-models with complex homogeneous target spaces and their deformations. These models are classically integrable~\cite{BykovNon, BykovSols, BykovZeroCurv, CYa}, at least in the sense that their e.o.m. admit a zero-curvature representation. This is a generalization and extension of the results about $\sigma$-models with symmetric target spaces (Grassmannians $G(m, n)$ for symmetry group $SU(n)$) that have been studied in the past, cf.~\cite{Adda1, Adda2, Eichenherr0, Eichenherr1, Din1, Din2, Perelomov0, Morozov1, Perelomov1}. Complex homogeneous spaces are torus bundles over flag manifolds~\cite{Wang}, and in the present paper we will for simplicity restrict to flag manifolds themselves.

\vspace{0.3cm}\noindent
It turns out that not only the homogeneous models are integrable, but also their deformations of a special kind. In the language of integrable systems, homogeneous models correspond to the `rational' case, and the deformations correspond to the trigonometric and elliptic cases. At the level of the $\sigma$-model action these deformations are rather involved, and the history of constructing deformed actions is rather long. Even the literature dedicated to the deformed principal chiral model  is vast, so we mention an indicative list of papers \cite{Cherednik, Fateev1, Klimcik1, Klimcik2, Klimcik3, Lukyanov, DelducHoare}. The method was subsequently generalized to the case of symmetric target spaces in~\cite{DMVq}.

\subsection[Hermitian deformations]{Hermitian deformations.}
An important advance has been achieved recently in the work~\cite{CYa} (building on earlier work~\cite{CYa1, CYa2}), where a general approach to deformations of $\sigma$-models has been developed. Within this approach one manifestly introduces a classical $r$-matrix of the corresponding type (rational/trigonometric/elliptic) in the action of the model. Apart from this conceptual advantage, the method applies to a wider class of models, in particular to the models with complex homogeneous target spaces proposed in~\cite{BykovNon, BykovSols, BykovZeroCurv}. These target spaces are not symmetric in general, but their deformations can now also be constructed. In the present paper our main goal is to study deformations of flag manifold $\sigma$-models in the so-called $\beta\gamma$-formulation, which will be reviewed in sec.~\ref{coupledbetagamma} below. The salient feature of this deformation is that it explicitly preserves the Hermiticity of the metric. This is in fact different from the deformations of symmetric spaces constructed in~\cite{DMVq}, i.e. deformations based on the $\mathbb{Z}_2$-grading of the Lie algebra of the symmetry group. In that case, as was shown on the example of the $\CP^{n-1}$-model in~\cite{DemulderCP}, the target space is generalized K\"ahler (see also~\cite{BykovLust} for an explanation of the different types of deformations and~\cite{LitvinovCP, FateevCP} for a study of a Toda-like field theory, `dual' to the deformed $\sigma$-model). Even if the undeformed geometry is K\"ahler (as in the case of Grassmannians $G(m, n)$), the Hermitian deformation that we consider in the present paper does not preserve the K\"ahler property. In fact for general flag manifolds even the homogeneous metric  is not K\"ahler.

\subsection[The Ricci flow conjecture]{The Ricci flow conjecture.}
In the special case of $\CP^1$ the deformation is essentially unique -- this is the so-called `sausage' deformation constructed long ago in~\cite{Onofri}\footnote{In a recent paper~\cite{BykovLust} we presented a derivation of the sausage model from the perspective of Pohlmeyer reduction~\cite{Pohlmeyer}.}. Already back then it was observed that the deformed metric is an explicit solution to the Ricci flow equation (see~\cite{Perelman, Hamilton1, Hamilton2} for a mathematical introduction to Ricci flow and a discussion of the history of the subject), which arises in this context as an equation of renormalization group flow driven by the $\beta$-function of the theory. More precisely, if appropriate coordinates are used, the only parameter that flows with `RG-time' $\tau$ is the deformation parameter itself. Moreover, this flow is extremely simple: if one denotes the trigonometric (multiplicative) deformation parameter $s$, then $s=e^{2\tau}$. We will call this flow linear, since $\mathrm{log}(s)\sim \tau$. It is reasonable to ask, whether this is a common feature of trigonometric integrable models. Up to our knowledge, in all known cases indeed the deformed geometry is stable under the flow, but the dependence of the deformation on $\tau$ might be rather complicated (as discussed in sec.~\ref{fateevsec} below, it is already non-linear in the case of Fateev's flow~\cite{Fateev1, BakasKong} that describes a deformation of $S^3$; see also~\cite{Lukyanov} for more general solutions in the case of $S^3$). One of the conjectures in~\cite{CYa} is that for Hermitian deformations the flow is again linear\footnote{In~\cite{BykovLust} we found that the flow is also linear in the case of generalized K\"ahler deformation of $\CP^{n-1}$ -- this is the deformation that arises from the construction of~\cite{DMVq}.}. Whether this is true is one of the key questions that we seek an answer to in the present paper. Our conclusion is that the conjecture does hold, however the Ricci flow equation has to be seen in an extended framework, when one includes the $B$-field, dilaton $\Phi$ and a compensating connection that we call $\mathcal{E}$. These ingredients acquire a transparent meaning from the standpoint of $\beta\gamma$-systems.  

\subsection[Affine and projective $\beta\gamma$-systems]{Affine and projective $\beta\gamma$-systems.}

$\beta\gamma$-systems (see~\cite{Witten02, NekrasovBeta} for an introduction) are interesting theories in their own right, but for the purposes of the present paper they will mostly serve as a way of writing $\sigma$-model Lagrangians in a first-order formalism of sorts. The most basic model is the one with target space $\CC^n$ and a particular metric, which in the undeformed case is really the standard metric on $\mathbb{R}\times S^{2n-1}$. We refer to this model (or similar models with target space $\CC^n\otimes \CC^m$) as the \emph{affine} model. In section~\ref{betagammaUG} we show that this model is a deformed extension of a Thirring-type model in bosonic incarnation, and compute its one-loop $\beta$-function. We then demonstrate that the Ricci flow equation has the linear solution $\mathrm{log}(s)=n\tau$, once one passes to certain natural coordinates. \emph{Projective} models may be obtained from affine ones by gauging part of the global symmetry. Such are the $\CP^{n-1}$ $\sigma$-model, Grassmannian and flag manifold models. It turns out that the symmetry that one needs to gauge is a chiral symmetry. As a result, the procedure is generally invalidated due to anomalies. Fortunately, these anomalies can be cancelled by coupling the theory minimally to fermions~(more complicated couplings with interacting fermions could also be possible). A striking property of this coupling is that, at the same time, it cancels the anomalies in L\"uscher's nonlocal charge~\cite{Luscher} (which arise in the $\CP^{n-1}$~\cite{Abdalla1} or Grassmannian~\cite{Abdalla2} models) and renders these theories integrable at the quantum level (those results are also collected in the book~\cite{AbdallaBook}). We conjecture that this is a general property, and that the theories so obtained -- $\sigma$-models in the $\beta\gamma$ formulation (Thirring-like models) coupled minimally to fermions -- are integrable at the quantum level. Evidence in favor of this conjecture is presented in section~\ref{integrconj}.

\vspace{0.3cm}\noindent
\emph{Notation.} Throughout the paper we will assume that the worldsheet is the complex plane $\CC$, with coordinates $z, \bar{z}$. Derivatives with respect to these coordinates will be denoted $\dd:=\dd_z$ and $\bd:=\dd_{\bar{z}}$. Similar notation is adopted for covariant derivatives, whose meaning is explained later on, i.e. $D, \bar{D}, \mathscr{D}, \bar{\mathscr{D}}$. For any matrix $Y$ the bar $\bar{Y}:=Y^\dagger$ will mean Hermitian conjugation.

\section{The classical $r$-matrix and coupled $\beta\gamma$-systems}\label{coupledbetagamma}

We start by introducing the models of study -- the coupled $\beta\gamma$-systems -- and explaining their classical integrability properties. This requires the notion of a classical $r$-matrix.

\subsection[The classical $r$-matrix]{The classical $r$-matrix.}\label{CYBEsec}

In its conventional form, the classical Yang-Baxter equation (CYBE) involves the classical $r$-matrix $r(u)\in \mathfrak{g}\otimes\mathfrak{g}$, where $\mathfrak{g}$ is a semi-simple or, more generally, reductive Lie algebra. $u$ is a parameter taking values in a complex abelian group. The equation itself takes values in $\mathfrak{g}\otimes\mathfrak{g}\otimes\mathfrak{g}$ and has the following form:
\bea
[r_{12}(u), r_{13}(u\cdot v)]+[r_{12}(u), r_{23}(v)]+[r_{13}(u\cdot v), r_{23}(v)]=0\,.
\eea
Since we mostly have the trigonometric case in mind, we write the equation in multiplicative form, that is to say $u, v \in \CC^\times$. The notation $r_{12}(u)$ means $r_{12}(u)=r(u)\otimes \mathds{1}$, and analogously for other pairs of indices. Solutions to the above equation have been extensively studied in the classical paper~\cite{BD}.

\vspace{0.3cm}\noindent
For the purposes of the present paper it is more convenient to think of the $r$-matrix as a map $r(u): \mathfrak{g} \to \mathfrak{g}$, or equivalently $r(u)\in \mathrm{End}(\mathfrak{g})\simeq \mathfrak{g}\otimes \mathfrak{g}^\ast$. In this case we will write $r_u(a)\in \mathfrak{g}$ for the $r$-matrix acting on a Lie algebra element $a\in \mathfrak{g}$. One also often assumes the so-called `unitarity' property of the $r$-matrix: 
\bea\label{rmunit}
\mathrm{Tr}(r_u(A)\,B)=-\mathrm{Tr}(A\,r_{u^{-1}}(B))\,.
\eea
As we will see shortly, for our purposes it will be useful to weaken this condition slightly. In the new notations the CYBE looks as follows:
\bea
[r_u(a), r_{uv}(b)]+r_u([r_v(b), a])+r_{uv}([b, r_{v^{-1}}(a)])=0\,.
\eea

\vspace{0.3cm}\noindent
The solution of interest has the form (for now we assume $\mathfrak{g}\simeq \mathfrak{su}_n$)
\bear\label{trigrmatrix}
&&r_u=\upalpha_u\,\pi_++\upbeta_u \,\pi_-+\upgamma_u\,\pi_0\,,\\ \label{abg}
&&\upalpha_u=\frac{u}{1-u},\quad\quad \upbeta_u=\frac{1}{1-u},\quad\quad \upgamma_u={1\over 2}\,\frac{1+u}{1-u}\,,
\eear
where $\pi_\pm$ are projections on the upper/lower-triangular matrices, and $\pi_0$ is the projection on the diagonal. The rational limit is achieved by setting $u=e^{-\upepsilon}$ and taking the limit $\upepsilon\to 0$, in which case $r_u\to \frac{1}{\upepsilon}\,\mathrm{Id}$.

\subsubsection[The constant $\mathcal{R}$-matrix]{The constant $\mathcal{R}$-matrix.}

The ansatz
\begin{empheq}[box=\fbox]{align}
\hspace{1em}\vspace{1em} \label{rR}
r_u=\frac{1}{2}\,\frac{1+u}{1-u}\,\mathrm{Id}+{i\over 2}\,\mathcal{R}\quad
\end{empheq}
transforms the CYBE to an equation on $\mathcal{R}$, which does not depend on the spectral parameter:
\bea\label{mCYBE}
[\Rc(a), \Rc(b)]+\Rc([\Rc(b), a]+[b, \Rc(a)])-[a, b]=0\,.
\eea
It is known in the literature as the `classical modified Yang-Baxter equation'. The solution~(\ref{trigrmatrix}) corresponds to\footnote{Another option is taking an $\mathcal{R}$-matrix induced by a complex structure on the Lie group $G$ with Lie algebra $\mathfrak{g}$~\cite{BykovCompDef}. In this case~(\ref{mCYBE}) is the condition of vanishing of the Nijenhuis tensor.} $\mathcal{R}=i\,(\pi_+-\pi_-)$.

\subsection[Classically integrable models of coupled $\beta\gamma$-systems]{Classically integrable models of coupled $\beta\gamma$-systems.}\label{classintsec}

The paper~\cite{CYa} introduced a method for constructing classically integrable models, including deformations of $\sigma$-models on homogeneous spaces, using classical $r$-matrices. The homogeneous models correspond to the rational case, and below we will mostly be interested in trigonometric deformations. In~\cite{BykovNilp} we proposed gauged linear $\sigma$-model (GLSM) formulations for these systems.  Introducing three matrices $U \in \mathrm{Hom}(\CC^{m}, \CC^n), V\in \mathrm{Hom}(\CC^n, \CC^{m}), \Phi\in \mathrm{End}(\CC^{n})$ (we will always assume $m\leq n$), we write down the Lagrangian\footnote{The Boltzmann weight in the path integral is $e^{-{1\over \hbar}\int\,d^2z\, \mathrsfso{L}}$, so that the kinetic term in~$\mathrsfso{L}$ is imaginary, and the potential term is real and positive (for $\hbar>0$), as  it should be in the case of first order actions in Euclidean space, cf.~\cite{ZJ}.}: 
\bea\label{lagr5}
\mathrsfso{L}=\mathrm{Tr}\left(V \bar{\mathscr{D}} U\right)-\mathrm{Tr}\left(V \bar{\mathscr{D}} U\right)^\dagger+\mathrm{Tr}\left(r_s^{-1}(\Phi) \bar{\Phi} \right)\,,
\eea
where $r_s$ is the classical $r$-matrix, depending on the deformation parameter $s$, that we encountered in section~\ref{CYBEsec}. The covariant derivative is ${\bar{\mathscr{D}}U=\bd U+i\,\bar{\Phi} U+i \,U \bar{\mathcal{A}}}$, where $\mathcal{A}$ is a gauge field, whose structure depends on the actual target space under consideration, and $\Phi, \bar{\Phi}$ are auxiliary fields\footnote{From the perspective of~\cite{CYa}, $\Phi, \bar{\Phi}$ are the components of the four-dimensional Chern-Simons gauge field along the worldsheet.}. One can eliminate these fields,  since they enter the Lagrangian quadratically. As a result, one arrives at the following:
\begin{empheq}[box=\fbox]{align}\label{lagr6}
\hspace{1em}\vspace{1em}
\mathrsfso{L}=\mathrm{Tr}\left(V \bar{D} U\right)-\mathrm{Tr}\left(V \bar{D} U\right)^\dagger+\mathrm{Tr}\left(r_s(U V) (U V)^\dagger\right)\,,\quad
\end{empheq}
where $\bar{D}U=\bd U+i \,U \bar{\mathcal{A}}$. This Lagrangian is the main object of study in the present paper.

\subsubsection[The zero curvature representation]{The zero curvature representation.}

The main property of the system~(\ref{lagr5})-(\ref{lagr6}) is that its e.o.m. admit a zero-curvature representation. To write  it down, we observe that in the undeformed case, when $r_s$ is proportional to the identity operator, the above Lagrangians have an $SU(n)$ global symmetry and a corresponding Noether current one-form
$
J=J\,dz+\bar{J}\,d\bar{z}=UV\,dz+\bar{V} \,\bar{U}\,d\bar{z}\,.
$
Using this one-form, we define a family of connections, following~\cite{CYa}: 
\bea
\mathscr{A}=r_{\kappa_1}(J)\,dz-r_{\kappa_2}(\bar{J})\,d\bar{z}\,.
\eea
Here $\kappa_1, \kappa_2$ are complex parameters that will be related below. We wish to prove that the connection $\mathscr{A}$ is flat:
\bea\label{flatA}
d\mathscr{A}+\mathscr{A}\wedge \mathscr{A}=-dz\wedge d\bar{z}\,\left(r_{\kappa_2}(\dd \bar{J})+r_{\kappa_1}(\bd J)+[r_{\kappa_1}(J), r_{\kappa_2}(\bar{J})]\right)\overset{?}{=}0\,.
\eea
To this end, we will use the equations of motion of the model~(\ref{lagr6}). To write them out, we define the `conjugate' operator $\hat{r}$ by the relation $\mathrm{Tr}(r_s(A)\,B)=-\mathrm{Tr}(A\,\hat{r}_{s^{-1}}(B))$. When the unitarity relation~(\ref{rmunit}) holds, $\hat{r}=r$. The e.o.m. take the form
\bear\label{Veq1}
&&\!\!\!\!\!\!\!\!\!\!\!\!\!\!\!\!\!\!\!\!V:\quad\quad \bar{D}U-\hat{r}_{s^{-1}}(\bar{V}\, \bar{U})\,U=0 \\ \label{Veq2}
&&\!\!\!\!\!\!\!\!\!\!\!\!\!\!\!\!\!\!\!\!\bar{V}:\quad\quad D \bar{U}-\bar{U}\,r_s(U V)=0 \\ 
&&\!\!\!\!\!\!\!\!\!\!\!\!\!\!\!\!\!\!\!\!U:\quad\quad \bar{D}V+V\,\hat{r}_{s^{-1}}(\bar{V}\,  \bar{U})=0 \\  
&&\!\!\!\!\!\!\!\!\!\!\!\!\!\!\!\!\!\!\!\!\bar{U}:\quad\quad D \bar{V}+r_s(U V)\,\bar{V}=0\,.
\eear
This implies the following concise equations for the `Noether current' $J$:
\bear
&&\bd J=[\hat{r}_{s^{-1}}(\bar{J}), J]\,,\quad\quad
\dd \bar{J}=[\bar{J}, r_s(J)]\,.
\eear
Substituting in the equation~(\ref{flatA}), we see that it is satisfied if the matrix $r$ obeys the equation
\bea\label{CYBE1}
r_{\kappa_2}([\bar{J}, r_s(J)])+r_{\kappa_1}([\hat{r}_{s^{-1}}(\bar{J}), J])+[r_{\kappa_1}(J), r_{\kappa_2}(\bar{J})]=0\,.
\eea
The reason why in our case $\hat{r}$ is not necessarily equal to $r$ is that we will mostly be dealing with a non-simple Lie algebra $\mathfrak{gl}_n=\mathfrak{sl}_n\oplus \CC$ ($\CC\mysub \mathfrak{gl}_n$ corresponds to matrices proportional to the unit matrix). We will assume a block-diagonal $r$-matrix, acting as follows:
\bea\label{glnr}
r_s=(r_s)_{\mathfrak{sl}_n}+(r_s)_{\CC}\,,\quad\quad (r_s)_{\mathfrak{sl}_n}\in \mathrm{End}(\mathfrak{sl}_n),\quad\quad (r_s)_{\CC}:=b(s)\,\mathrm{Tr}.
\eea
Here $(r_s)_{\mathfrak{sl}_n}$ acts on traceless matrices as in~(\ref{rR}), and $(r_s)_{\CC}$ acts on matrices of the type $\alpha\cdot\mathds{1}$ as multiplication by  $n \,b(s)$. In this case $\hat{r}_s=r_s-(b(s^{-1})+b(s))\,\mathrm{Tr}$. If the unitarity relation is satisfied, the mismatch vanishes. However, in either case $b(s)$ completely drops out from the equation~(\ref{CYBE1}), so we will prefer allowing an arbitrary function $b(s)$ for the moment. Postulating the relations $\kappa_1=u,\; \kappa_2=uv,\; v=s^{-1}$ (implying $\kappa_1=\kappa_2\,s\,$) between the parameters, we identify~(\ref{CYBE1}) with the classical Yang-Baxter equation for $\mathfrak{g}=\mathfrak{sl}_n$ from section~\ref{CYBEsec}. 

\section{Affine $\beta\gamma$-systems and their deformations}\label{betagammaUG}

We start from the discussion of arguably simplest $\beta\gamma$-systems -- the ones with affine target space $\CC^n$, as well as related $\sigma$-models\footnote{In fact, in the context of $\beta\gamma$-systems it is somewhat more convenient to talk about complex `phase space', which in this case is~$T^\ast \CC^n$.}. From the standpoint of the general system~(\ref{lagr6}) this is the situation when the gauge field vanishes: $\mathcal{A}=0$. 

\subsection[The $\CC^n$-model in the $\beta\gamma$-formulation: the bosonic Thirring model]{The $\CC^n$-model in the $\beta\gamma$-formulation: the bosonic Thirring model.}\label{Thirringsec}

Our first object of study will be a $\sigma$-model with target space that is formally $\CC^n$, written as a coupling of two $\beta\gamma$-systems. We introduce the coordinates $U \in \mathrm{Hom}(\CC, \CC^n), V\in \mathrm{Hom}(\CC^n, \CC)$ on $T^\ast \CC^n$ and consider the Lagrangian~(\ref{lagr6}) with $\mathcal{A}=0$: 
\bea\label{lagr11}
\mathrsfso{L}=V \cdot \bd U-\dd\bar{U} \cdot \bar{V}+\mathrm{Tr}\left(r_s(U V) (U V)^\dagger\right)\,,
\eea
In the undeformed case, when $r_s=\frac{\mathds{1}}{\upvarepsilon}+\left( b-{1\over n\upvarepsilon}\right)\,\mathrm{Tr}$, one can easily integrate out the $V$-variables to arrive at the conventional form of the $\sigma$-model Lagrangian
\bea\label{undeflagr}
\mathrsfso{L}=\frac{\upvarepsilon}{\bar{U}U}\cdot\bd U\cdot \left(\mathds{1}+\updelta\,\frac{\bar{U}\otimes U}{\bar{U}U}\right)\cdot\dd\bar{U}\,,\quad\quad \updelta={1\!-\!b\upepsilon n\over n\!-\!1\!+\!b\upepsilon n}\,.
\eea
This model is representative of the full class of models we will be considering. Apart from a Hermitian metric, it features a $B$-field of a very particular type -- it is the fundamental Hermitian form of the metric. In other words, if $\cal{J}$ is the complex structure, with respect to which the metric is Hermitian, the $B$-field is $B=g\circ \cal{J}$. Models with this property appeared in~\cite{WittenTop}, and gauged Wess-Zumino-Novikov-Witten theories with this feature were studied in~\cite{Gawedzki, ShataBoer}.

\vspace{0.3cm}\noindent
To give the reader a spirit of the target spaces that we will be dealing with, let us set $n=1$ for the moment, in which case the line element is $ds^2={1\over b}\,\frac{ d U d\bar{U}}{ U \bar{U}}$. This is the standard flat metric on a cylinder $\CC^\times$ with multiplicative coordinate $U$. Although the formal target space might have been $\CC$, the form of the interactions is such that the origin is effectively excluded, so that the actual target space is $\CC^\times$. This is a general feature: for arbitrary $n$ the target space is $\CC^n\setminus\{0\}$, due to the presence of the denominators in~(\ref{undeflagr}).

\vspace{0.3cm}\noindent
A very fruitful way of looking at the system~(\ref{lagr11}) comes from rewriting it in Dirac form. We introduce $n$ `Dirac bosons'
\bea
\Psi_a=\begin{pmatrix} U_a\\ \bar{V}_a\end{pmatrix}\,,\quad\quad a=1, \ldots, n\,.
\eea
Sometimes we will suppress the $SU(n)$ index, in which case $\Psi\in \mathrm{Hom}(\CC, \CC^2\otimes \CC^n)$.
The Lagrangian~(\ref{lagr11}) is, in these notations (and upon integrating by parts),
\begin{empheq}[box=\fbox]{align}\label{psilagr}
\hspace{1em}\vspace{1em}
\mathrsfso{L}=\bar{\Psi_a} \slashed{\dd} \Psi_a + (r_s)^{cd}_{ab}\,\left(\bar{\Psi_a}{1+\gamma_5\over 2}\Psi_c\right) \,\left(\bar{\Psi_d}{1-\gamma_5\over 2}\Psi_b\right)\,.\quad
\end{empheq}
The Dirac notations are standard: $\sigma_{1, 2}$ are the Pauli matrices,  $\slashed{\dd}:=\sum\limits_{i=1}^2\, \sigma_i \,\dd_i$ and $\gamma_5:=i\,\sigma_1\sigma_2$. If one sets $(r_s)^{cd}_{ab}\sim \delta_a^c \delta_b^d$, the system becomes the bosonic incarnation of the so-called chiral Gross-Neveu model (equivalently the $SU(n)$ Thirring model~\cite{WittenThirring}). For ${n=1}$ one obtains the bosonic version of the Thirring model. As is known from the fermionic theory, the $\beta$-function of the Thirring model vanishes, which is compatible with the fact that the bosonic version is a $\sigma$-model with target space the flat cylinder~$\CC^\times$.

\vspace{0.3cm}\noindent
The model~(\ref{psilagr}) is `chiral',  meaning that there is a symmetry\footnote{In Minkowski signature this would have been the usual $U(1)$ chiral symmetry. This difference in chiral transformations has been observed in~\cite{Zumino, Mehta}.}
\bea
U\to \uplambda U, \quad V\to \uplambda^{-1} V\,, \quad\quad \textrm{where}\quad\quad \uplambda\in \CC^\times\,.
\eea
A general $SU(n)$-invariant Lagrangian would only retain a $U(1)$-symmetry, $|\uplambda|=1$, and is not invariant under the full $\CC^\times$, which arises in~(\ref{psilagr}) due to the chiral projectors. In other words, for Euclidean worldsheet signature chiral symmetry is equivalent to the complexification of the original (non-chiral) symmetry. This chiral symmetry is of extreme importance at least for two reasons:
\begin{itemize}
\item Chiral symmetry ensures that the quartic interaction terms are quadratic in the $V$-variables, which are the `momenta' conjugate to $U$. This allows integrating them out and arriving at a metric form of the $\sigma$-model.
\item We will also be interested in $\sigma$-models on projective spaces, Grasmannians etc., and these may be obtained by taking the quotient w.r.t. $\CC^\times$, i.e. by gauging the chiral symmetry. In doing so, one needs to verify that it is free of anomalies.
\end{itemize}

\subsection[The $\beta$-function and the Ricci flow]{The $\beta$-function and the Ricci flow.}\label{betafuncsec}
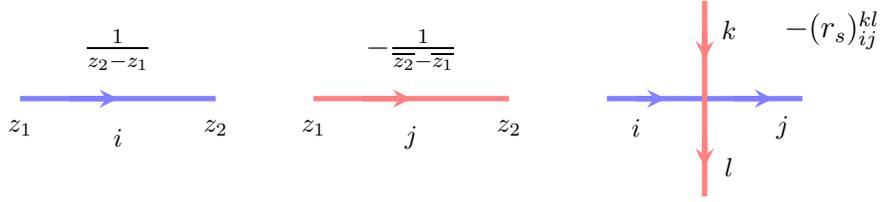
\begin{figure}
\centering
\bea\nonumber
\begin{tikzpicture}[
baseline=-\the\dimexpr\fontdimen22\textfont2\relax,scale=1.3]
\draw[-stealth, blue!50, line width=2pt, rounded corners] (-1,0)  -- (0,0) node[right,black] {};
\draw[blue!50, line width=2pt, rounded corners] (-0.5,0)  -- (1,0) node[right,black] {};
\draw[-stealth, red!50, line width=2pt, rounded corners] (2,0)  -- (3,0) node[right,black] {};
\draw[red!50, line width=2pt, rounded corners] (2.5,0)  -- (4,0) node[right,black] {};
\draw[-stealth, blue!50, line width=2pt, rounded corners] (5,0)  -- (5.6,0) node[right,black] {};
\draw[-stealth, blue!50, line width=2pt, rounded corners] (5.3,0)  -- (6.7,0) node[right,black] {};
\draw[blue!50, line width=2pt, rounded corners] (6.3,0)  -- (7,0) node[right,black] {};
\draw[-stealth, red!50, line width=2pt, rounded corners] (6,1)  -- (6,0.4) node[right,black] {};
\draw[-stealth, red!50, line width=2pt, rounded corners] (6,0.7)  -- (6,-0.7) node[right,black] {};
\draw[red!50, line width=2pt, rounded corners] (6,-0.4)  -- (6,-1) node[right,black] {};
\node at (-1,-0.3) {\footnotesize $z_1$};
\node at (1,-0.3) {\footnotesize  $z_2$};
\node at (0,0.5) {${1\over z_2-z_1}$};
\node at (0,-0.4) {\footnotesize $i$};
\node at (2,-0.3) {\footnotesize  $z_1$};
\node at (4,-0.3) {\footnotesize  $z_2$};
\node at (3,0.5) {$-{1\over \bar{z_2}-\bar{z_1}}$};
\node at (3,-0.4) {\footnotesize $j$};
\node at (5.3,-0.3) {\footnotesize $i$};
\node at (6.8,-0.3) {\footnotesize $j$};
\node at (6.25,0.7) {\footnotesize $k$};
\node at (6.25,-0.7) {\footnotesize $l$};
\node at (7.3,0.7) { $-(r_s)_{ij}^{kl}$};
\end{tikzpicture}
\eea
\caption{Feynman rules of the deformed bosonic Thirring model~(\ref{lagr11}).} \label{fig1}
\end{figure}

\begin{figure}
\centering
\bea\nonumber
\begin{tikzpicture}[
baseline=-\the\dimexpr\fontdimen22\textfont2\relax,scale=1.2]
\draw [-stealth, red!50, line width=2pt] (2,0) arc [radius=1, start angle=100, end angle= 120];
\draw [-stealth, red!50, line width=2pt] (2,0) arc [radius=1, start angle=100, end angle= 180];
\draw [-stealth, red!50, line width=2pt] (2,0) arc [radius=1, start angle=100, end angle= 255];
\draw [red!50, line width=2pt] (2,0) arc [radius=1, start angle=100, end angle= 260];
\draw [-stealth, blue!50, line width=2pt] (1,0) arc [radius=1, start angle=80, end angle= 60];
\draw [-stealth, blue!50, line width=2pt] (1,0) arc [radius=1, start angle=80, end angle= 0];
\draw [-stealth, blue!50, line width=2pt] (1,0) arc [radius=1, start angle=80, end angle= -75];
\draw [blue!50, line width=2pt] (1,0) arc [radius=1, start angle=80, end angle= -80];
\node at (2.2,0) {\footnotesize $k$};
\node at (0.8,0) {\footnotesize $i$};
\node at (0.8,-2) {\footnotesize $j$};
\node at (2.2,-2) {\footnotesize $l$};
\draw [-stealth, red!50, line width=2pt] (5,-1.95) arc [radius=1, start angle=260, end angle= 240];
\draw [-stealth, red!50, line width=2pt] (5,-1.95) arc [radius=1, start angle=260, end angle= 180];
\draw [-stealth, red!50, line width=2pt] (5,-1.95) arc [radius=1, start angle=260, end angle= 110];
\draw [red!50, line width=2pt] (5,-1.95) arc [radius=1, start angle=260, end angle= 100];
\draw [-stealth, blue!50, line width=2pt] (4,0) arc [radius=1, start angle=80, end angle= 60];
\draw [-stealth, blue!50, line width=2pt] (4,0) arc [radius=1, start angle=80, end angle= 0];
\draw [-stealth, blue!50, line width=2pt] (4,0) arc [radius=1, start angle=80, end angle= -75];
\draw [blue!50, line width=2pt] (4,0) arc [radius=1, start angle=80, end angle= -80];
\node at (5.2,0) {\footnotesize $l$};
\node at (3.8,0) {\footnotesize $i$};
\node at (3.8,-2) {\footnotesize $j$};
\node at (5.2,-2) {\footnotesize $k$};
\end{tikzpicture}
\eea
\caption{Diagrams contributing to the $\beta$-function at one loop.} \label{fig2}
\end{figure}
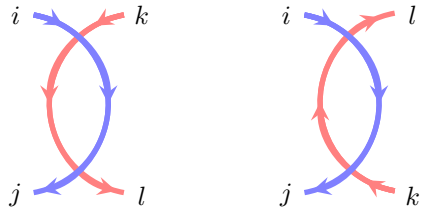

Having discussed the analogy with the chiral Thirring model, we return to the Lagrangian~(\ref{lagr11}). The main question we pose is whether this Lagrangian preserves its form after renormalization, at least to one loop order -- in other words, whether it is sufficient to renormalize the parameters of the $r$-matrix. To this end we write out the Feynman rules of the system in Fig.~\ref{fig1}.
At one loop the two diagrams contributing to the renormalization of the quartic vertex are shown in Fig.~\ref{fig2}.
There is a relative sign between the two diagrams, due to the different directions of the lines in the loops. Otherwise, the type of the divergence is the same -- it is logarithmic\footnote{Here we are talking about UV divergences.}, proportional to~$\int\frac{d^2z}{z \bar{z}}$. As a result, the one-loop $\beta$-function is

\begin{empheq}[box=\fbox]{align}\label{betafunc}
\hspace{1em}\vspace{1em}
\beta_{ij}^{kl}=\sum\limits_{p, q=1}^{n}\,\left((r_s)_{ip}^{kq} (r_s)_{pj}^{ql}-(r_s)_{ip}^{ql} (r_s)_{pj}^{kq}\right)\quad
\end{empheq}
As already discussed earlier, we will assume a block-diagonal $r$-matrix~(\ref{glnr}), where $(r_s)_{\mathfrak{sl}_n}$ acts as in~(\ref{rR}), and $(r_s)_{\CC}$ acts on a unit matrix as multiplication by $n\cdot b(s)$. This is translated into the four-index notation as follows:
\bea\label{rsu1}
(r_s)_{ij}^{kl}=\underbracket[0.6pt][0.6ex]{a_{kl}(s)\,\left(\delta_i^k \delta_j^l-{1\over n}\delta_{ij}\delta^{kl}\right)}_{:=(r^{\mathrm{T}}_s)_{ij}^{kl}}+b(s)\,\delta_{ij}\delta^{kl}\,.
\eea
We will refer to the piece proportional to $b$ as the \emph{longitudinal mode}, and the $\mathfrak{su}_n$-piece will be called the \emph{transverse} $r^{\mathrm{T}}$-matrix . Later on, in the discussion of gauged models, we will describe a more transparent interpretation of the longitudinal term. The coefficients $a_{kl}(s)$ are defined by the action of~(\ref{rR}) in the standard basis: $r_s(e_k\otimes e_l)=a_{kl}(s)\,e_k\otimes e_l$. Concretely (see~(\ref{abg})),
\bea\label{acoefs}
a_{ij}=\begin{cases}
\quad \frac{s}{1-s}=\upalpha,\quad\quad \;\;i<j\\
\quad\frac{1}{1-s}=\upbeta,\quad\quad \;\;i>j\\
\quad{1\over 2}\frac{1+s}{1-s}=\upgamma,\quad\quad i=j\,.
\end{cases}
\eea
Substituting~(\ref{rsu1}) in~(\ref{betafunc}), we obtain
\bea
\beta_{ij}^{kl}=\delta_i^k \delta_j^l\,\left(\sum\limits_{p=1}^n\,a_{ip}a_{pj}\right)-\delta_{ij}\delta^{kl}\,a_{ik}a_{ki}
\eea
A direct calculation shows that
\bea
a_{ij}a_{ji}=\frac{s}{(1-s)^2}+{1\over 4}\delta_{ij}, \quad\quad
\sum\limits_{p=1}^n\,a_{ip}a_{pj}={1\over 4}\delta_{ij}+\frac{ns}{(1-s)^2}+(i-j)\,a_{ij}
\eea
Returning to the $\beta$-function,
\bea\label{betafunc2}
\beta_{ij}^{kl}=\left[\frac{ns}{(1-s)^2}+(i-j)a_{ij}\right]\,\underbracket[0.6pt][0.6ex]{\left(\delta_i^k \delta_j^l-{1\over n}\delta_{ij}\delta^{kl}\right)}_{:=\Pi_{ij}^{kl}}\,.
\eea
Since $a_{ij}={1\over 2}\frac{1+s}{1-s}+{i\over 2}\,\mathcal{R}_{ij}$, the one-loop result is not of the form~(\ref{rsu1}) (due to the term proportional to $i-j$). For this reason the straightforward Ricci flow equation ${d r_s\over d\tau}=\beta$ for $s(\tau)$ does not have a solution. However, this can be easily remedied by allowing reparametrizations of coordinates along the flow. Let us reparametrize
\bea\label{rescaling}
U\to D U, \quad V\to V D^{-1}\,,\quad\quad \textrm{where} \quad\quad D=\mathrm{Diag}\{\kappa_1, \ldots , \kappa_n\}\,.
\eea
The kinetic term in~(\ref{lagr11}) is invariant, so the only effect is in the effective replacement of the $r$-matrix by $\tilde{r}$, where $(\tilde{r}_s)_{ij}^{kl}=\frac{\kappa_i \bar{\kappa_k}}{\kappa_j \bar{\kappa_l}}(r_s)_{ij}^{kl}$. One has to conjugate the $\beta$-function tensor analogously, and the equation we will aim to solve is ${d \tilde{r}_s\over d\tau}=\tilde{\beta}$. It may be rewritten as an equation for the original $r$-matrix as follows:
\bear
&&{d\over d\tau}(r_s)_{ij}^{kl}=\beta_{ij}^{kl}-{d\over d\tau}\left(\log{\left(\frac{\kappa_i \bar{\kappa_k}}{\kappa_j \bar{\kappa_l}}\right)}\right)\,(r_s)_{ij}^{kl}:=\widehat{\beta}_{ij}^{kl}\,\\ \nonumber
&&\textrm{where}\quad\quad \widehat{\beta}_{ij}^{kl}=\left[\frac{ns}{(1-s)^2}+(i-j)a_{ij}-{d\over d\tau}\left(\log{\left(\big|{\kappa_i\over\kappa_j}\big|^2\right)}\right) a_{ij}\right]\,\Pi_{ij}^{kl}
\eear
The unwanted term may now be canceled by the simple substitution
\bea\label{coordflow}
\kappa_j=e^{{\tau\over 2}\,j}\,.
\eea
Recalling again the expression for $a_{ij}(s)$, we find that the remaining equations may be written as $\dot{b}=0$, ${d\over d\tau}\left({1\over 2}{1+s\over 1-s}\right)={ns\over (1-s)^2}$, or $\dot{s}=ns$. Therefore
\begin{empheq}[box=\fbox]{align}\label{sflow}
\hspace{1em}
b=\mathrm{const.},\quad s=e^{n\,\tau}\quad 
\end{empheq}
We arrive at the important conclusion that the trigonometrically deformed system~(\ref{lagr11}) provides a solution to the Ricci flow equation.

\subsubsection[Endpoints of the Ricci flow]{Endpoints of the Ricci flow.}

Let us return to the geometric picture and analyze the effect of the flow~(\ref{coordflow})-(\ref{sflow}) on the metric of the model. For $\tau\to 0$ one has $s\to 1$, so that, after integrating out $V, \bar{V}$, one is left with the Lagrangian~(\ref{undeflagr}), where $\upvarepsilon\simeq 1-s$. This is the undeformed model with vanishing radius, i.e. a growing coupling constant. Physically this is the IR limit of the model. The opposite (UV) regime is $\tau\to-\infty$ (this is a so-called `ancient' solution -- the one defined on a half-line -- in the terminology adopted in the literature on Ricci flow), in which case $s\to 0$. One should not forget, however, that we performed a rescaling of coordinates~(\ref{rescaling}) by the $\kappa$-factors~(\ref{coordflow}), so that in the rescaled coordinates the interaction terms in the Lagrangian take the form
\bear\nonumber
&&\mathrm{Tr}\left(r_s(U V) (U V)^\dagger\right)
=\sum\limits_{i, j=1}^n\,a_{ij}\,{|\kappa_i|^2\over |\kappa_j|^2}|U_i|^2 |V_j|^2+\left(b-{\upgamma(\tau)\over n}\right)\big|\sum\limits_{i=1}^n\,U_i V_i\big|^2=\\&&=\sum\limits_{i, j=1}^n\,a_{ij}\,e^{(i-j)\,\tau}|U_i|^2 |V_j|^2+\left(b-{\upgamma(\tau)\over n}\right)\big|\sum\limits_{i=1}^n\,U_i V_i\big|^2\,.
\eear

\vspace{-0.1cm}\noindent
From the explicit form~(\ref{acoefs}) of the $r$-matrix coefficients $a_{ij}$ it is easy to see that in the limit $\tau\to -\infty$ only the diagonal terms survive: $\underset{\tau\to -\infty}{\mathrm{lim}}\,(a_{ij}\,e^{(i-j)\,\tau})={1\over 2}\,\delta_{ij}$. Besides, $\underset{\tau\to -\infty}{\mathrm{lim}}\, \upgamma(\tau)={1\over 2}$. After integrating out $V, \bar{V}$ the limiting Lagrangian takes the form 
\bea
\mathrsfso{L}\big|_{\tau\to -\infty}=\sum\limits_{i=1}^n\,\frac{\bd U_i \dd\bar{U}_i}{U_i \bar{U}_i}-\frac{1}{n+{1\over 2}\upkappa_0}\,\big|\sum\limits_{i=1}^n\,{\bd U_i\over U_i}\big|^2,\quad\quad \upkappa_0=\left(b-{1\over 2n}\right)^{-1}\,.
\eea
Clearly this is a flat metric on an $n$-dimensional cylinder $(\CC^\times)^n$, with multiplicative coordinates $U_1, \ldots, U_n$. We conclude that the Ricci flow interpolates between the homogeneous system~(\ref{undeflagr}) (with vanishing overall scale) and a flat $n$-dimensional cylinder. The latter could be called an `asymptotically free' regime.

\vspace{0.3cm}\noindent
We can make this consideration slightly more precise, using the notion of $\mathbb{Z}_n$-grading on the Lie algebra $\mathfrak{gl}_n$. We will be using the standard grading that schematically may be depicted as
\bea\footnotesize
\parenMatrixstack{
0 & \textbf{1} & 2 & 3 & \ldots & n\!-\!1\\
n\!-\!1& 0 & \textbf{1} & 2 & \ldots & n\!-\!2\\
n\!-\!2\,\,& n\!-\!1 & 0 & \textbf{1} & \ldots & n\!-\!3\\
\vdots & \vdots & \vdots & \vdots & \ddots & \vdots\\
\textbf{1} & 2 &3 &4 & \ldots& 0
}
\eea
In fact, the grading is completely determined by the grading of the positive simple roots and the `imaginary root' (cf.~\cite{BykovAnomaly} for a recent discussion), which is why we have emphasized those in bold. The decomposition of the Lie algebra is $\mathfrak{gl}_n=\oplus_{i=0}^{n-1} \mathfrak{g}_i$ ($[\mathfrak{g}_i, \mathfrak{g}_j]\mysub \mathfrak{g}_{i+j\,\mathrm{mod}\, n}$), and the projection on the component $\mathfrak{g}_i$ will be denoted $\pi_i$. Since $D=\mathrm{Diag}\{e^{j\tau\over 2}\}_{j=1}^n$ and recalling that $s=e^{n\tau}$, one finds that the matrix entering the quartic interaction is really

\vspace{-0.3cm}
\bea\label{Znproj}
\mathrm{Ad}_D r_s \mathrm{Ad}_D={1\over 2}\,\pi_0+\frac{1}{1-e^{n\tau}}\sum\limits_{j=1}^{n}\,e^{j\tau}\,\pi_{n-j}\,.
\eea
The sum vanishes in the limit $\tau\to-\infty$, so that $\mathrm{Ad}_D r_s \mathrm{Ad}_D(UV)\to {1\over 2}\pi_0(UV)$.

\subsection[$\CC^n\otimes \CC^m$ model]{$\CC^n\otimes \CC^m$ model.}

The setup of the previous sections allows a simple generalization, when, instead of taking $U$ and $V$ as vectors, we may take them as matrices of size $m\times n$, as in sec.~\ref{classintsec}, that is to say  $U \in \mathrm{Hom}(\CC^m, \CC^n), V\in \mathrm{Hom}(\CC^n, \CC^m)$. The Lagrangian has the same form as before, up to an obvious inclusion of traces in suitable places:
\bea\label{lagr12}
\mathrsfso{L}=\mathrm{Tr}(V \cdot \bd U)-\mathrm{Tr}(\dd\bar{U} \cdot \bar{V})+\mathrm{Tr}\left(r_s(U V) (U V)^\dagger\right)\,,
\eea
The Feynman rules, shown in Fig.~\ref{fig3}, include an additional index $\alpha, \beta, \ldots \in \{1\ldots m\}$ that we denote by a Greek letter to distinguish from the $SU(n)$ indices written in Latin.
\begin{figure}
\centering
\bea
\begin{tikzpicture}[
baseline=-\the\dimexpr\fontdimen22\textfont2\relax,scale=1.3]
\draw[-stealth, blue!50, line width=2pt, rounded corners] (-1,0)  -- (0,0) node[right,black] {};
\draw[blue!50, line width=2pt, rounded corners] (-0.5,0)  -- (1,0) node[right,black] {};
\draw[-stealth, red!50, line width=2pt, rounded corners] (2,0)  -- (3,0) node[right,black] {};
\draw[red!50, line width=2pt, rounded corners] (2.5,0)  -- (4,0) node[right,black] {};
\draw[-stealth, blue!50, line width=2pt, rounded corners] (5,0)  -- (5.6,0) node[right,black] {};
\draw[-stealth, blue!50, line width=2pt, rounded corners] (5.3,0)  -- (6.7,0) node[right,black] {};
\draw[blue!50, line width=2pt, rounded corners] (6.3,0)  -- (7,0) node[right,black] {};
\draw[-stealth, red!50, line width=2pt, rounded corners] (6,1)  -- (6,0.4) node[right,black] {};
\draw[-stealth, red!50, line width=2pt, rounded corners] (6,0.7)  -- (6,-0.7) node[right,black] {};
\draw[red!50, line width=2pt, rounded corners] (6,-0.4)  -- (6,-1) node[right,black] {};
\node at (-1,-0.3) {\footnotesize $z_1$};
\node at (1,-0.3) {\footnotesize  $z_2$};
\node at (0,0.5) {${1\over z_2-z_1}$};
\node at (0,-0.4) {\footnotesize $i, \alpha$};
\node at (2,-0.3) {\footnotesize  $z_1$};
\node at (4,-0.3) {\footnotesize  $z_2$};
\node at (3,0.5) {$-{1\over \bar{z_2}-\bar{z_1}}$};
\node at (3,-0.4) {\footnotesize $j, \alpha$};
\node at (5.3,-0.3) {\footnotesize $i, \alpha$};
\node at (6.8,-0.3) {\footnotesize $j, \alpha$};
\node at (6.35,0.8) {\footnotesize $k, \beta$};
\node at (6.35,-0.85) {\footnotesize $l, \beta$};
\node at (7.4,0.7) { $-(r_s)_{ij}^{kl}$};
\end{tikzpicture}
\eea
\caption{Feynman rules of the $\CC^n\otimes \CC^m$ model. Latin and Greek indices correspond to $SU(n)$ and $SU(m)$, respectively.} \label{fig3}
\end{figure}
Notice that the Greek indices are preserved along the blue and red lines separately. As a result, they are `spectator' indices for the divergent diagrams in Fig.~\ref{fig2}, and the $\beta$-function is the same as before, given by~(\ref{betafunc}). In other words, the $\beta$-function is independent of~$m$. We will give a geometric interpretation of this fact~in~Sec.~\ref{sumrulesec}.

\subsection[The geometric approach]{The geometric approach.}\label{geomapprsec}

In section~\ref{Thirringsec} we performed a direct calculation of the $\beta$-function of the model. We could have followed another route: first integrating out the $V, \bar{V}$-fields, arriving at a $\sigma$-model with target space $\CC^n$, computing its metric, $B$-field and dilaton and checking that they satisfy the generalized Ricci flow equations\footnote{For the $\beta$-functions in the r.h.s. of these equations see~\cite{Curci, Polchinski}.}
\begin{empheq}[box=\fbox]{align}
\hspace{1em}\vspace{1em}
\label{Ricflowaff}
&-\dot{g_{ij}}=R_{ij}+{1\over 4}H_{imn}H_{jm'n'}g^{mm'}g^{nn'}+2\,\nabla_i\nabla_j \Phi\,,\quad\\ \nonumber
&-\dot{B_{ij}}=-{1\over 2}\,\nabla^k\,H_{kij}+\nabla^k\Phi\,H_{kij}\,,\quad\\ \nonumber
&-\dot{\Phi}={\mathrm{const.}}-{1\over 2}\,\nabla^k\nabla_k\,\Phi+\nabla^k\Phi \nabla_k \Phi+{1\over 24}\,H_{kmn}H^{kmn}\,,
\end{empheq}
where $H=dB$ is the field strength tensor of the $B$-field, and $\Phi$ is the dilaton. Realizing this program is a substantially more complicated procedure. A direct calculation is possible for $n=2$, and some indications in this direction are presented in Appendix~\ref{appRicci} (otherwise we have used numerical calculations to check the statement for low values of $n$).  Here we will only write out the explicit expressions for the metric, $B$-field and dilaton for arbitrary~$n$ ($\upkappa=(b-{\upgamma\over n})^{-1}$):
\bear\label{Cnmetricdef1}
&&\hspace{-2cm}ds^2=
\sum\limits_{i, j=1}^n\,k_{ij} (dt_i\,dt_j+d\phi_i d\phi_j)\,, \quad\quad B=
\sum\limits_{i, j=1}^n\,k_{ij}\, dt_i\wedge d\phi_j\,,\\ \nonumber && \\ \nonumber &&\hspace{-3cm} \quad\quad \textrm{where}\hspace{2cm}   {1\over 2}\,k_{ij}=\lambda_i\,\delta_{ij}-\frac{\lambda_i \lambda_j}{\upkappa+\sum\limits_k \lambda_k}\,,\\ \nonumber &&
\hspace{0.9cm} \lambda_m^{-1}=\upalpha\,\sum_{j<m}\,e^{2\,(t_j-t_m)}+\upbeta\,\sum_{j>m}\,\,e^{2\,(t_j-t_m)}+\upgamma\,,\\ \nonumber
&&\hspace{0.9cm} e^{-2\Phi}=\prod\limits_{k=1}^n\,{e^{2\,t_k}\over \lambda_k}\times \left(1+{1\over \upkappa}\,\sum\limits_{k=1}^n \lambda_k \right)\,.
\eear
The dilaton $\Phi$ appears from the determinant arising upon integration over the $V$-variables, in exactly the same way as it happens in the context of Buscher rules for $T$-duality~\cite{Buscher, TseytlinDil, SchwarzDil}.

\subsubsection[Comparison with Fateev's flow]{Comparison with Fateev's flow.}\label{fateevsec}

The fact that the one-dimensional space of metrics, with parameter $s$, is closed under Ricci flow is rather remarkable. What is even more exceptional for the models considered in the present paper is that the evolution of $s$ is particularly simple: $s=e^{n\tau}$, see~(\ref{sflow}). To illustrate this point, let us consider another basic example of integrable model -- the $\sigma$-model with target space the squashed sphere $S^3_\eta$. This is a one-parametric sub-family of a two-parameter family of integrable deformations of $S^3$ that were constructed in~\cite{Fateev1, Klimcik1}. The one-parametric deformation is defined by the Lagrangian
\bea
\mathrsfso{L}=\mathrm{Tr}\left(J^\dagger {1\over r_u}\,J\right),\quad\quad J=g^\dagger dg\,, \quad\quad g\in SU(2)\,.
\eea
Parametrizing $g=\begin{pmatrix} Z_1 & - \bar{Z}_2 \\ Z_2 & \bar{Z}_1\end{pmatrix}$ and $Z_1=\cos{({\theta\over 2})}\,e^{i(\psi+\phi)}, Z_2=\sin{({\theta\over 2})}\,e^{i\psi}$, we find the expression for the deformed metric
\bea
ds^2=\underbracket[0.6pt][0.6ex]{\;{2\over \upgamma}\;}_{:=f}\,\left(d\psi+\cos^2{\theta\over 2}\,d\phi\right)^2+{1\over 4}\underbracket[0.6pt][0.6ex]{\left({1\over \upalpha}+{1\over \upbeta}\right)}_{:=g}\left(d\theta^2+\sin^2{\theta} \,d\phi^2\right)
\eea
$\upalpha, \upbeta, \upgamma$ are the same parameters as before, see~(\ref{acoefs}), but with $s$ replaced by $u$. The Ricci tensor of this metric has the same form, but with the replacements $f\to f_1=4(1-{g\over 2f})$, $g\to g_1={2 g^2\over f}$. The Ricci flow equation $-\dot{g_{ij}}=R_{ij}$ is equivalent to $\dot{f}=-f_1, \dot{g}=-g_1$, and upon substituting the expressions of $f, g, f_1, g_1$ in terms of the deformation parameter $u$ one obtains the equation
\bea
\dot{u}={4 u^2\over (1+u)^2}\,.
\eea
In terms of $u=e^{\eta}$ the solution is ${1\over 2}(\eta+\mathrm{sinh}\, \eta)=\tau$. This is the result found in~\cite{Fateev1}. As we see, the dependence of $\eta$ on $\tau$ is nonlinear, in contrast to the situation we encountered earlier.

\vspace{0.3cm}\noindent
We note in passing that there is a beautiful generalization~\cite{BakasKong} of Fateev's result for more general Hopf-like fibrations of homogeneous spaces ${G\over K} \to {G\over H}$ with fiber $H\over K$ in the case when all three of these spaces admit Einstein metrics $\mathds{G}_E^{G\over K}, \mathds{G}_E^{G\over H}, \mathds{G}_E^{H\over K}$, and a certain inequality on the cosmological constants is satisfied. Then, if the fiber is not flat, there is a second Einstein metric on the total space $G\over K$, and a solution of the Ricci flow equation of the type $\mathds{G}=g(\tau)\,\mathds{G}_E^{G\over H}+f(\tau)\,\mathds{G}_E^{H\over K}$, interpolating between the two Einstein metrics on $G\over K$. If the fiber is flat, the behavior is as in  the Hopf fibration example above: for $\tau\to 0$ one recovers the round metric on the total space and in the limit $\tau\to -\infty$ the size of the base blows up, whereas the size of the fiber remains finite.

\section{Projective (gauged) $\beta\gamma$-systems and their deformations}\label{betagammaProj}

In the previous section we considered $\beta\gamma$-systems, and related $\sigma$-models, with affine target spaces. Ultimately we are interested in $\sigma$-models with projective target spaces, such as Grassmannians and flag manifolds. This is done by gauging a part of the symmetry group of the theory -- a process that is typically violated by chiral anomalies, as we will now explain.

\subsection[The $\CP^{n-1}$-model]{The $\CP^{n-1}$-model.}\label{cpnsec}

We start with the simplest example: the complex projective space. It can be obtained from the affine space $\CC^n$ by gauging the $\CC^\times$ symmetry group acting as $U_i \to \uplambda U_i$, $V_i \to \uplambda^{-1} V_i$. As explained in section~\ref{Thirringsec}, the case $|\uplambda|=1$ corresponds to the usual $U(1)$ symmetry of the theory (charge conservation), however for $\uplambda \in \CC^\times$ this is a chiral transformation: the radial $\mathbb{R}^\times \mysub \CC^\times$ is what would have been the chiral $U(1)$ in Minkowski signature. Such chiral transformations are typically subject to anomalies. The latter can be deduced from Schwinger's calculation of the effective action of the gauge fields $\mathcal{A}$, which comes from the one-loop determinant of the matter fields:
\bea\label{effaction}
\mathrsfso{S}_{\mathrm{eff.}}={\upxi\over 2} \int\,dz\,d\bar{z}\,F_{z\bar{z}}{1\over \triangle}F_{z\bar{z}}\,,\quad\quad F_{z\bar{z}}=i\,(\dd \bar{\mathcal{A}}-\bd \mathcal{A})\,.
\eea
The coefficient $\upxi$ collects some numerical factors and is proportional to the number of matter fields we have integrated over. The action is invariant w.r.t. the gauge transformations of the original $U(1)$, $\mathcal{A}\to \mathcal{A}+\dd \alpha$,  $\bar{\mathcal{A}}\to \bar{\mathcal{A}}+\bd \alpha$ ($\alpha\in \mathbb{R}$), but not w.r.t. the complexified gauge transformations $\mathcal{A}\to \mathcal{A}+\dd \alpha$,  $\bar{\mathcal{A}}\to \bar{\mathcal{A}}+\bd \bar{\alpha}$ ($\alpha\in \CC$)\footnote{In fact, there is no local function of the fields $\mathcal{A}, \bar{\mathcal{A}}$ and their derivatives that would be invariant w.r.t. this transformation.}.

\vspace{0.3cm}\noindent
The anomaly also manifests itself in another way. This perspective will shed some light on the parameter $b$ of the `longitudinal mode' of the $r$-matrix~(\ref{rsu1}). The crucial observation is that the deformed metric on $\CP^{n-1}$ may be obtained by taking a formal limit $b\to \infty$~($\upkappa\to 0$) of the $\CC^n$-metric in~(\ref{Cnmetricdef1}): one can check, for example, that the matrix $k_{ij}$ in~(\ref{Cnmetricdef1}) is degenerate in this limit, pointing at an additional local projective invariance. Why this happens becomes clear, if we rewrite the $\CC^n$-system with the $r$-matrix~(\ref{rsu1}) as follows:
\bea\label{stuckgauge}
\mathrsfso{L}=V \cdot \bar{D} U-D\bar{U} \cdot \bar{V}+\mathrm{Tr}\left(r^{\mathrm{T}}_s(U V) (U V)^\dagger\right)+{1\over b}\,|\mathcal{A}|^2\,,\quad\quad \bar{D} U=\bd U +i\,U \bar{\mathcal{A}}.
\eea
This step is just a Hubbard-Stratonovich transformation applied to the longitudinal term $b\,|V\cdot U|^2$ in the original Lagrangian. Notice that in the quartic interaction terms one now has the $\mathfrak{su}_n$ $r$-matrix $r^{\mathrm{T}}$. If one takes the formal limit $b\to \infty$, the system~(\ref{stuckgauge}) becomes the $\beta\gamma$-system of the projective space $\CP^{n-1}$. An important  subtlety, however, is that there is a non-zero one-loop correction to the coefficient $1\over b$ in front of $|\mathcal{A}|^2$ (cf.~\cite{ModInv} for a calculation in Euclidean signature). This correction is the same coefficient $\upxi$ that features in the effective action~(\ref{effaction}). Indeed, the $\mathcal{A}\bar{\mathcal{A}}$-part of the effective action is $\upxi \int\,dz\,d\bar{z}\,\dd \bar{\mathcal{A}}{1\over \triangle}\bd \mathcal{A}=-\upxi \int\,dz\,d\bar{z}\,|\mathcal{A}|^2$, where we have integrated by parts. As a result, effectively in the Lagrangian~(\ref{stuckgauge}) one has to make the shift
\begin{empheq}[box=\fbox]{align}
\hspace{1em}\vspace{1em} \label{bshift}
{1\over b}\to {1\over b}-\upxi\,.
\end{empheq}
This shift invalidates the conclusion that in the $b\to \infty$ limit one gets the $\beta\gamma$-system of~$\CP^{n-1}$. In order to be able to gauge the complexified symmetry group $\CC^\times$, one has to cancel this anomaly. A natural way to do it is to add fermionic partners of the bosonic fields $U, V$. As the bosonic model may be written in the form~(\ref{psilagr}) with the help of a `Dirac boson' $\Psi$, the fermionic extension should take the form\footnote{See~\cite{Waldron} for a review of spinors in Euclidean spaces.} 
\bear\label{cpnferm}
&&\mathrsfso{L}=\bar{\Psi_a} \slashed{D} \Psi_a + (r_s^{\mathrm{T}})^{cd}_{ab}\,\left(\bar{\Psi_a}{1+\gamma_5\over 2}\Psi_c\right) \,\left(\bar{\Psi_d}{1-\gamma_5\over 2}\Psi_b\right)+\bar{\Theta_a} \slashed{D} \Theta_a\,,\\ \nonumber && \Psi, \Theta\in \mathrm{Hom}(\CC,  \CC^2\otimes\CC^n)\,.
\eear
Here $\Theta_a, \,a=1, \ldots, n$ are $n$ Dirac fermions. For such a theory the one-loop anomalies coming from integrations over bosons and fermions exactly cancel each other, i.e. the coefficient $\upxi$ in the shift~(\ref{bshift}) is zero, so that one can take the limit $b\to\infty$.

\subsubsection[Relation to the conventional form of the $\CP^{n-1}$-model]{Relation to the conventional form of the $\CP^{n-1}$-model.}

The model~(\ref{cpnferm}) just introduced is classically equivalent to the usual $\CP^{n-1}$-model minimally interacting with $n$ Dirac fermions. To see this, we again pass to the $(U, \bar{V})$-parametrization of the field $\Psi$ and integrate out the fields $V$ that enter quadratically:
\bea
\mathrsfso{L}=\upvarepsilon\cdot \frac{D\bar{U} \,\bar{D} U }{\bar{U}U}+ \bar{\Theta} \slashed{D} \Theta\,.
\eea
Since there is no chiral anomaly, we impose the gauge $\bar{U}U=1$, in which case 
\bea\label{standcpn}
\mathrsfso{L}={\upvarepsilon\over 2}\cdot h^{\alpha\beta}\, D_\alpha\bar{U} D_\beta U+ \bar{\Theta} \slashed{D} \Theta\,+{1\over 2}\epsilon_{\alpha\beta}\,\dd_\alpha \bar{U}\dd_\beta U 
\eea
$h$ is the flat metric on the worldsheet, and $\epsilon$ is the standard skew-symmetric matrix. The last term in~(\ref{standcpn}) is topological, and the first two terms correspond to the standard $\CP^{n-1}$-model interacting with $n$ Dirac fermions $\Theta$.

\vspace{0.3cm}\noindent
There is a similar relation between $\sigma$-models in the $\beta\gamma$-formulation and standard $\sigma$-models when the target space is symmetric, i.e. a Grassmannian. In the general case of flag manifold models one should start from the $\beta\gamma$-formulation, and a rewriting in the form~(\ref{standcpn}) would exhibit a non-topological (non-closed) $B$-field.

\subsubsection[Ricci flow for the Hermitian $\CP^{n-1}$ model]{Ricci flow for the Hermitian $\CP^{n-1}$ model.}\label{cpnriccisec}

Just as in the case of the affine model we expect the Ricci flow equations to be satisfied. We find this to be the case, yet with the addition of Lie derivatives w.r.t. a certain new field $\mathcal{E}$ (which we interchangeably view as a vector or one-form): 
\begin{empheq}[box=\fbox]{align}
\hspace{1em}\vspace{1em}
\label{Ricflowproj}
&-\dot{g_{ij}}=R_{ij}+{1\over 4}H_{imn}H_{jm'n'}g^{mm'}g^{nn'}+2\,\nabla_i\nabla_j \Phi+(\pounds_{n \mathcal{E}} \,g)_{ij}\,,\\ \nonumber
&-\dot{B_{ij}}=-{1\over 2}\,\nabla^k\,H_{kij}+\nabla^k\Phi\,H_{kij}+
(\pounds_{n \mathcal{E}} \,B)_{ij}\,,\quad\\ \nonumber
&-\dot{\Phi}={\mathrm{const.}}-{1\over 2}\,\nabla^k\nabla_k\,\Phi+\nabla^k\Phi \nabla_k \Phi
+{1\over 24}\,H_{kmn}H^{kmn}+\pounds_{n \mathcal{E}} \,\Phi\,,\\ \nonumber
&\quad\quad\quad \textrm{where}\quad\quad\mathcal{E}=\mathrm{Re}(A),\quad\quad A=\frac{\sum\limits_{m=1}^n\hat{\lambda}_m\,\bar{\hat{U}}_m d\hat{U}_m}{\sum\limits_{m=1}^n\hat{\lambda}_m\,|\hat{U}_m|^2}\,.
\end{empheq}
Here $\hat{\lambda}_j=\lambda_j\,e^{-2t_j}$ and $\hat{U}$ are the inhomogeneous coordinates, obtained by setting $\hat{U}_n=1$. One of the effects of adding Lie derivatives is in replacing $d\Phi\to \mathcal{D}\Phi=d \Phi+n\,\mathcal{E}$. This has the following meaning. The dilaton of the affine model (see~(\ref{Cnmetricdef1})) is not invariant under the global $\CC^\times$ symmetry (it is shifted by a constant instead). Therefore while gauging the symmetry one needs to make the derivative $d\Phi$ covariant, and this is exactly what the field $\mathcal{E}$ achieves. In the undeformed case, when $s\to 1$, one has $\Phi\to -{n\over 2}\,\log{(\sum\limits_{i=1}^n\,|U_i|^2)}$ and $\mathcal{E}\to {1\over 2}\,d\log{(\sum\limits_{i=1}^n\,|U_i|^2)}$, so that $\mathcal{D}\Phi\to 0$, and the dilaton effectively disappears from the first two equations.

\vspace{0.3cm}\noindent
It is useful to look at the above equations~(\ref{Ricflowproj}) from another perspective, namely as a limit~$b\to\infty$ of the corresponding equations~(\ref{Ricflowaff}) for target space $\CC^n$, since, as we discussed at the beginning of sec.~\ref{cpnsec}, the $\CP^{n-1}$-model may be obtained from the $\CC^n$-model in this limit. Most terms converge to the corresponding ones of $\CP^{n-1}$, and complications come from terms involving the dilaton. In the first equation this is the term $\nabla_i\nabla_j \Phi$. To take the limit we first rewrite the metric~(\ref{Cnmetricdef1}) of the deformed~$\CC^n$ at large~$b$. It is useful to parametrize $U_i=e^p\,\hat{U}_i$, where $\hat{U}_i$ are the inhomogeneous coordinates on $\CP^{n-1}$, so for example $\hat{U}_n=1$, and $p$ is the additive coordinate in the $\CC^\times$-fiber. Then
\bea\label{blimmetr}
ds^2\simeq {2\over b}\left(|dp|^2+d\bar{p} A+dp \bar{A}\right)+[ds^2]_{\mathrm{T}}\,,\quad\quad A=\frac{\sum\limits_{m=1}^n\hat{\lambda}_m\,\bar{\hat{U}}_m d\hat{U}_m}{\sum\limits_{m=1}^n\hat{\lambda}_m\,|\hat{U}_m|^2}\,.
\eea
Here $[ds^2]_{\mathrm{T}}$ is the `transverse' metric, which converges to the $\CP^{n-1}$ metric in the limit $b\to \infty$. It is clear from the expression~(\ref{Cnmetricdef1}) for the dilaton that it is in fact linear in $\mathrm{Re}(p):=t$, i.e. $\Phi=-nt+\ldots$, where the ellipsis denotes terms independent of~$t$. Let $i, j, k, l$ be `transverse' indices, corresponding to the $\CP^{n-1}$ directions, parametrized by $\hat{U}$. Then $\nabla_i\nabla_j \Phi=\nabla_i^{\mathrm{T}}\nabla_j^{\mathrm{T}} \Phi+n\Gamma^t_{ij}$, where the superscript $t$ corresponds to the `radial' $t$-direction, and $\nabla_i^{\mathrm{T}}$ are the covariant derivatives w.r.t. the transverse metric. The geometric meaning of $\Gamma^t_{ij}$ is that it is part of the second fundamental form of the transverse manifold sitting in the ambient space at a fixed value of $p$. What remains is to compute the relevant Christoffel symbols for the metric~(\ref{blimmetr}): $\Gamma^t_{ij}={1\over 2} g^{tt}\left({\dd g_{ti}\over \dd x^j}+{\dd g_{tj}\over \dd x^i}\right)+g^{tk} \Gamma_{k|ij}$. Since $g^{tA}g_{A l}=\delta^t_l=0$ and $g^{tA}g_{A l}=g^{tt}g_{t l}+g^{tk}g_{k l}$ (summation over $A$ is w.r.t. all directions), we find that in the limit $g^{tk}\simeq -g^{tt}g_{t l} g^{lk}$. Since $g_{t l}={1\over b} (A_l+\bar{A}_l)={2\over b} \mathcal{E}_l$ and $g^{tt}\simeq {b\over 2}$, we get $\Gamma^t_{ij}\simeq {1\over 2}(\nabla_i \mathcal{E}_j+\nabla_j \mathcal{E}_i)$, so that
\bea
2\,\nabla_i\nabla_j \Phi\simeq 2\,\nabla_i^{\mathrm{T}}\nabla_j^{\mathrm{T}}\Phi+n\,(\nabla_i \mathcal{E}_j+\nabla_j \mathcal{E}_i)=\nabla_i^{\mathrm{T}}\mathcal{D}_j^{\mathrm{T}} \Phi+\nabla_j^{\mathrm{T}}\mathcal{D}_i^{\mathrm{T}} \Phi\,
\eea
which is the expression from~(\ref{Ricflowproj}), once the superscript $\mathrm{T}$ is dropped.

\vspace{0.3cm}\noindent
Let us also derive the second equation in~(\ref{Ricflowproj}). We write $\nabla^A \,\Phi H_{Aij}=-n\,(g^{tt}\,H_{tij}+g^{tk}\,H_{kij})+\nabla^k \,\Phi\, H_{kij}=-n\,g^{tt}\,H_{tij}+\mathcal{D}^k \,\Phi\, H_{kij}$. In order to evaluate the first term in the limit $b\to\infty$, we write out the relevant terms in the three-form $H$ derived from the metric~(\ref{blimmetr}): $H=[dB]_{\mathrm{T}}+{1\over b}\,dt\wedge d\left({\bar{A}-A\over 2i}\right)+\ldots$ As a result, $g^{tt}\,H_{tij}\to \dd_i \mathcal{F}_j-\dd_j \mathcal{F}_i $, where $\mathcal{F}=\mathrm{Im}(A)$. To compare with the second equation in~(\ref{Ricflowproj}), we invoke the definition $(\pounds_{n \mathcal{E}} \,B)=n(i_{\mathcal{E}}dB+d(i_{\mathcal{E}}B))$. The first term is $n\, \mathcal{E}^k H_{kij}$, so it contributes to $\mathcal{D}^k\Phi\, H_{kij}$. To write out the second term we recall that $B=g\circ\mathcal{J}$, so that $(i_{\mathcal{E}}B)_i=\mathcal{E}_k\,g^{km}\,B_{mi}=\mathcal{J}_i^k\,\mathcal{E}_k$. Since $\mathcal{E}$ is the real part of a form of type $(1, 0)$, $\mathcal{J}\circ \mathcal{E}$ is its imaginary part: $i_{\mathcal{E}}B=\mathrm{Im}(A)=\mathcal{F}$. This completes the derivation of the equation for $\dot{B_{ij}}$ in~(\ref{Ricflowproj}).

\vspace{0.3cm}\noindent
Finally, in the equation for the dilaton the nontrivial term is $\nabla^k\Phi \nabla_k \Phi$. Its reduction from the affine equation looks as follows: $\nabla^A \Phi\,\nabla_A \Phi=\nabla^m \Phi\,\nabla_m \Phi+2 \,g^{tm} \nabla_m\Phi\cdot (-n)+g^{tt} \cdot (-n)^2$. To get the correct expression for $g^{tt}$ to order $O(1)$, one needs to take into account the subleading contribution in $g_{tt}$: $g_{tt}={2\over b}-{2\mathcal{Q}\over b^2}$, where $\mathcal{Q}=-{\upgamma\over n}+{1\over \sum\limits_m\hat{\lambda}_m\,|\hat{U}_m|^2}$. In this case $g^{tt}\simeq {b\over 2}+ \mathcal{E}_m g^{mk}\mathcal{E}_k+{\mathcal{Q}\over 2}$. Using $g^{tm}\simeq -g^{mk}\,\mathcal{E}_k$, we obtain $\nabla^A \Phi\,\nabla_A \Phi\to \mathcal{D}^k\Phi \mathcal{D}_k \Phi+n^2 {\mathcal{Q}\over 2}+\mathrm{const.}$ All ingredients are manifestly gauge-invariant ($\CC^\times$-invariant). The equation for $\dot{\Phi}$ follows thanks to the identity $\nabla^k \mathcal{E}_k-2 \mathcal{E}^k\,\mathcal{D}_k\Phi-n\,\mathcal{Q}=0$ that we have verified explicitly. Presumably the latter should be viewed as a gauge condition.

\vspace{0.3cm}\noindent
We note that the appearance of terms $\nabla_i \mathcal{E}_j+\nabla_j \mathcal{E}_i$ has been observed in supergravity equations in a very similar situation -- when the dilaton is linear in a coordinate along an isometry direction of the metric~\cite{TseytlinWulff} (see also~\cite{ArutTseytlin}).

\subsubsection[The inhomogeneous gauge]{The inhomogeneous gauge.}

In the previous section we explained that the Hermitian metric, $B$-field and dilaton of the deformed $\CP^{n-1}$-model satisfy the generalized Ricci flow equations, which are the RG-flow equations at one-loop. In doing so, we made use of the $\beta$-function~(\ref{betafunc}). The $\beta$-function may also be computed directly for the $\CP^{n-1}$-model by passing to inhomogeneous coordinates, which is a way of fixing a $\CC^\times$-gauge explicitly. We will now describe how this gauge is fixed in practice. This will lead us to a curious observation, that the gauge-fixed model is a theory of fields with polynomial interactions.

\vspace{0.3cm}\noindent
To start with, we choose 
\bea\label{inhomgauge}
U_n=1\,.
\eea
Variation of the Lagrangian~(\ref{cpnferm}) w.r.t. $\bar{\mathcal{A}}$ gives $VU+\textrm{fermionic terms} =0$, so that 
\bea\label{Vinhom}
V_n=-\sum\limits_{k=1}^{n-1}\,V_k U_k+\textrm{fermionic terms}.
\eea
This completely removes the gauge field at the expense of modifying the Feynman rules of the theory. Dropping fermionic fields for the moment, we write the Lagrangian of the model~(\ref{lagr6}) in this gauge:
\bear\label{inhomgaugelagr}
&&\mathrsfso{L}=\sum\limits_{k=1}^{n-1}\,\left(V_k \bd U_k-\bar{V}_k \dd \bar{U}_k+\upbeta |V_k|^2\right)+\\ \nonumber
&&\underbracket[0.6pt][0.6ex]{+\sum\limits_{l, m=1}^{n-1}\,a_{lm}\,|U_l|^2 |V_m|^2+\upgamma\,\big|\sum\limits_{p=1}^{n-1}\,U_p V_p\big|^2}_{\textrm{quartic vertices}}+\underbracket[0.6pt][0.6ex]{\upalpha \,\left(\sum\limits_{k=1}^{n-1}\,|U_k|^2\right)\,\big|\sum\limits_{p=1}^{n-1}\,U_p V_p\big|^2 }_{\textrm{sextic vertices}}
\eear
We see that the propagators and vertices are modified, and on top of that sextic vertices have appeared:
\bea
\begin{tikzpicture}[
baseline=-\the\dimexpr\fontdimen22\textfont2\relax,scale=1.3]
\draw[-stealth, red!50, line width=2pt, rounded corners] (-1,0)  --  (-0.5,0) node[right,black] {};
\draw[red!50, line width=2pt, rounded corners] (-0.6,0)  --  (0,0) node[right,black] {};
\draw[blue!50, line width=2pt, rounded corners] (0,0)  --  (0.6,0) node[right,black] {};
\draw[-stealth, blue!50, line width=2pt, rounded corners] (1,0)  -- (0.5,0)   node[right,black] {};
\draw[-stealth, blue!50, line width=2pt, rounded corners] (-0.45,0.9)  --  (-0.2,0.4) node[right,black] {};
\draw[-stealth, blue!50, line width=2pt, rounded corners] (-0.25,0.5) -- (0.25,-0.5)    node[right,black] {};
\draw[blue!50, line width=2pt, rounded corners] (0.2,-0.4) -- (0.45,-0.9)    node[right,black] {};

\draw[-stealth, red!50, line width=2pt, rounded corners] (0.45,0.9)  --  (0.2,0.4) node[right,black] {};
\draw[-stealth, red!50, line width=2pt, rounded corners] (0.25,0.5) -- (-0.25,-0.5)    node[right,black] {};
\draw[red!50, line width=2pt, rounded corners] (-0.2,-0.4) -- (-0.45,-0.9)    node[right,black] {};

\node at (-1,-0.2) {\footnotesize $p$};
\node at (1,-0.2) {\footnotesize $p'$};
\node at (-0.6,0.85) {\footnotesize $q$};
\node at (0.6,0.85) {\footnotesize $r$};
\node at (0.6,-0.85) {\footnotesize $q'$};
\node at (-0.6,-0.85) {\footnotesize $r'$};

\node at (3,0) {\footnotesize $-\upalpha\,  \delta_{pp'} \delta_{qq'} \delta_{rr'}$};
\end{tikzpicture}
\eea
\noindent
A fascinating feature of the Lagrangian~(\ref{inhomgaugelagr}) is that its interaction terms are polynomial in the $(U, V)$-variables. In other words, instead of a nonlinear $\sigma$-model we have arrived at a different nonlinear theory -- the theory of several bosonic fields (albeit with fermionic propagators) and their fermionic partners with polynomial interactions.

\vspace{0.5cm}\noindent
The quadratic form in the above Lagrangian is $\begin{pmatrix} 0& \dd\\ \bd&-\upbeta\end{pmatrix}$. Its inverse is $\begin{pmatrix} {\upbeta \over \dd\bd}& \bd^{-1}\\ \dd^{-1}&0\end{pmatrix}$. This has the interesting consequence that the propagators $\langle U\,V\rangle$ and $\langle \bar{U}\,\bar{V}\rangle$ are not modified, but there appears an additional propagator
\bea
\begin{tikzpicture}[
baseline=-\the\dimexpr\fontdimen22\textfont2\relax,scale=1.3]
\draw[-stealth, blue!50, line width=2pt, rounded corners] (0,0)  --  (-0.5,0) node[right,black] {};
\draw[blue!50, line width=2pt, rounded corners] (-0.4,0)  --  (-1,0) node[right,black] {};
\draw[-stealth, red!50, line width=2pt, rounded corners] (0,0)  -- (0.5,0) node[right,black] {};
\draw[red!50, line width=2pt, rounded corners] (0.4,0)  -- (1,0) node[right,black] {};
\node at (-1,-0.3) {\footnotesize $z_1$};
\node at (1,-0.3) {\footnotesize  $z_2$};
\node at (0,-0.4) {\footnotesize $i$};
\node at (5, 0) {$\langle U_i(z_1)\,\bar{U}_j(z_2) \rangle\sim \delta_i^j\,\upbeta\log|z_1-z_2|^2\,,$};
\end{tikzpicture}
\eea
i.e. it looks like a propagator of a scalar field in two dimensions. As a result, at one loop there will be additional logarithmically divergent bubble-like diagrams. To illustrate how this works we compute in Appendix~\ref{betafuncinhomspp} the $\beta$-function in the inhomogeneous gauge~(\ref{inhomgauge}).

\subsection[The deformed flag manifold $\sigma$-model]{The deformed flag manifold $\sigma$-model.}\label{defflagsec}

So far we have considered Hermitian deformations of the $\CP^{n-1}$-model. Next we wish to show that the whole setup can be easily generalized to arbitrary flag manifolds
\bea\label{flaguni}
\mathscr{F}_{n_1, \ldots, n_s}:=\frac{U(n)}{U(n_1)\times \cdots \times U(n_s)}\,.
\eea
Since the $\beta\gamma$-systems are defined for complex manifolds, we will need an additional piece of data on $\mathscr{F}_{n_1, \ldots, n_s}$ -- the complex structure. Complex structures on $\mathscr{F}_{n_1, \ldots, n_s}$ are in 1-to-1 correspondence with the orderings of the integers $n_1, \ldots, n_s$~(cf.~\cite{BykovAnomaly}). For simplicity we will assume that the complex structure is specified by the ordering of the indices of $n_i$'s. In this case we may think of $\mathscr{F}_{n_1, \ldots, n_s}$ as the space of flags
\bear\label{flagseq}
&&0\mysub L_1 \mysub \ldots \mysub L_s\simeq \CC^n,\\
&&\textrm{where}\quad L_j\simeq \CC^{m_j} \quad \textrm{and} \quad m_j=\sum\limits_{k=1}^j\,n_k\,.
\eear
This gives rise to another definition of the flag manifold:
\bea\label{flagcomplex}
\mathscr{F}_{n_1, \ldots, n_s}\simeq \frac{GL(n, \CC)}{P_{m_1, \ldots, m_s}}\,,
\eea
where $P_{m_1, \ldots, m_s}$ is the parabolic subgroup of $GL(n, \CC)$ stabilizing flags of the form~(\ref{flagseq}).

\vspace{0.3cm}\noindent
The relevant gauged linear $\sigma$-model (GLSM) formulation for flag manifold models was developed in~\cite{BykovNilp}, where one took Nakajima's presentation of the flag manifold as an $A_{s-1}$ quiver variety as a starting point. First, one denotes the dimension of the maximal proper subspace in the flag~(\ref{flagseq}) as $m:=m_{s-1}=\sum\limits_{i=1}^{s-1}\,n_i$. Just as before, the Lagrangian is formulated in terms of three matrices $U\in \mathrm{Hom}(\CC^m, \CC^n), V\in \mathrm{Hom}(\CC^n, \CC^m), \Phi \in  \mathrm{End}(\CC^n)$ and has the following form:
\bea\label{lagrGL2}
\mathrsfso{L}= \mathrm{Tr}\left(V \bar{\mathscr{D}} U\right)-\mathrm{Tr}\left(V \bar{\mathscr{D}} U\right)^\dagger+\mathrm{Tr}\left((r_s^{\mathrm{T}})^{-1}(\Phi) \bar{\Phi}\right)\,,
\eea
The key fact about the above Lagrangian is that the gauge field $\mathcal{A}$ takes values in the Lie algebra $\mathfrak{p}=\mathrm{Lie}(P_{m_1, \ldots, m_{s-1}})$ of the parabolic subgroup of $GL(m, \CC)$. It is an upper-block-triangular matrix:

\vspace{-0.5cm}
\bea\label{colormatr}
\mathcal{A}=
\resizebox{5cm}{2.5cm}{%
\begin{tikzpicture}[
baseline=-\the\dimexpr\fontdimen22\textfont2\relax,scale=1]
\matrix[matrix of math nodes,left delimiter=(,right delimiter=),ampersand replacement=\&] (matr) {
\node (X0) {\ast};\&\ast\&\ast\&\ast\&\ast\&\node (X00) {\ast};\\
\node (X1) {\ast};\& \node (X2) {\ast};\&\ast\&\ast\&\ast\&\node (X11) {\ast};\\
\&\&\ast\&\ast\&\ast\&\ast\\
\&\&\node (X3) {\ast};\&\node (X4) {\ast};\&\ast\&\node (X22) {\ast};\\
\&\&\&\&\ast\&\ast\\
\&\&\&\&\node (X5) {\ast};\&\node (X6) {\ast};\\
};
\node[fill=blue!10, fit=(X0.north west) (X11.south east) ] {};
\node[fill=blue!10, fit=(X2.south east) (X22.south east) ] {};
\node[fill=blue!10, fit=(X4.south east) (X6.south east) ] {};
\draw[blue!50, line width=1pt, rounded corners] ([yshift=-3pt,xshift=-3pt]X1.south west)  -- ([yshift=-3pt,xshift=-2.5pt]X2.south east) node[right,black] {};
\draw[blue!50, line width=1pt, rounded corners] ([yshift=-3pt,xshift=-3pt]X2.south east)  -- ([yshift=-3pt,xshift=-3pt]X3.south west) node[right,black] {};
\draw[blue!50, line width=1pt, rounded corners] ([yshift=-3pt,xshift=-3pt]X3.south west)  -- ([yshift=-3pt,xshift=-2.5pt]X4.south east) node[right,black] {};
\draw[blue!50, line width=1pt, rounded corners] ([yshift=-3pt,xshift=-3pt]X4.south east)  -- ([yshift=-3pt,xshift=-3pt]X5.south west) node[right,black] {};
\draw[blue!50, dotted, line width=1pt] ([yshift=-3pt,xshift=-3pt]X3.south west)  -- ([yshift=-40pt,xshift=-3pt]X3.south west) node[right,black] {};
\draw[blue!50, dotted, line width=1pt] ([yshift=-3pt,xshift=-3pt]X5.south west)  -- ([yshift=-10pt,xshift=-3pt]X5.south west) node[right,black] {};
\draw[latex-latex] ([yshift=-6pt,xshift=-3pt]X5.south west) --  node[below, yshift=2pt] {{\scriptsize$ m_1$}} ([yshift=-6pt,xshift=3pt]X6.south east);
\draw[latex-latex] ([yshift=-40pt,xshift=-3pt]X3.south west) --  node[below, yshift=2pt] {{\scriptsize$ m_2$}} ([yshift=-40pt,xshift=55pt]X3.south west);
\draw[latex-latex] ([yshift=6pt,xshift=-3pt]X0.north west) --  node[above, yshift=-2pt] {{\scriptsize$ m$}} ([yshift=6pt,xshift=3pt]X00.north east);
\end{tikzpicture}
},\quad\quad\quad \bar{\mathcal{A}}=\mathcal{A}^\dagger\,.
\eea
At first sight it might look surprising that we are dealing with $m\times n$-matrices. There is in fact an alternative and (at least classically) equivalent formulation in terms of $n\times n$ matrices, as we explain in Appendix~\ref{defflagmetrapp}. The approach based on $m\times n$-matrices is better suited for the analysis of the $1\over n$-expansion~\cite{BykovAnomaly}, since in this setup one takes $n\to\infty$ while keeping $m$ fixed, so that the matrices $U, V$ are `vector-like', i.e. they grow in one direction. In Appendix~\ref{defflagmetrapp} we also provide a rewriting of the above Lagrangian in geometric ($\sigma$-model-like) form.

\subsubsection[Thirring-like form of the flag $\sigma$-model with fermions]{Thirring-like form of the flag $\sigma$-model with fermions.}\label{integrconj}

The pure bosonic flag manifold $\sigma$-model in the gauged $\beta\gamma$-formulation suffers from a chiral anomaly, in the same way as the $\CP^{n-1}$-model does. As before, it can be cancelled by adding fermions. For this reason we propose the following theory, which is free of chiral anomalies (the summation over the Greek indices is from $1$ to $m$, where $m=\sum\limits_{i=1}^{s-1}\,n_i$, and over the Latin indices from $1$ to $n$):
\begin{empheq}[box=\fbox]{align}
\hspace{1em}\vspace{1em}
\label{flaglagr}
&\mathrsfso{L}=\bar{\Psi_a^\alpha} \slashed{D} \Psi_a^\alpha + (r_s^{\mathrm{T}})^{cd}_{ab}\,\left(\bar{\Psi_a^\alpha}{1+\gamma_5\over 2}\Psi_c^\beta\right) \,\left(\bar{\Psi_d^\beta}{1-\gamma_5\over 2}\Psi_b^\alpha\right)+\bar{\Theta_a^\alpha} \slashed{D} \Theta_a^\alpha\,,\\ \nonumber & \Psi, \Theta\in \mathrm{Hom}(\CC^m,  \CC^n)\otimes \CC^2\,,\quad\quad
\Psi\;\;\textrm{bosonic},\;\; \Theta\;\; \textrm{fermionic}\,.
\end{empheq}
Here $\bar{D}^\alpha_\beta=\delta_\beta^\alpha\,\bd +i \, \bar{\mathcal{A}}^\alpha_\beta$ is the covariant derivative acting on the $GL(m, \CC)$ indices, and the gauge field $\mathcal{A}$ is of block-triangular form~(\ref{colormatr}). The fermionic determinant exactly cancels the bosonic one, hence the chiral anomalies are absent, and we may gauge the parabolic subgroup of $GL(m, \CC)$, as required.

\vspace{0.3cm}\noindent
We conjecture that the model~(\ref{flaglagr}) is quantum integrable. Evidence in favor of this conjecture is as follows:
\begin{itemize}
\item The bosonic part of the model is classically integrable~\cite{BykovNon,BykovSols, BykovZeroCurv, CYa}.
\item The cancellation of chiral anomalies is achieved by adding the fermionic fields $\Theta$. Incidentally, in the case of $\CP^{n-1}$ and Grassmannian models the same minimally coupled fermions remove the  anomaly in the non-local charge of the model~\cite{Abdalla1, Abdalla2}. This charge is directly related to the integrability of the model~\cite{Luscher}. In the pure bosonic flag manifold models the anomaly is certainly present~\cite{BykovAnomaly}. A feature of this anomaly is that, at least to leading order in the $1\over n$-expansion, it is of the same form as for the Grassmannian model, the only difference being the restricted form of the gauge field~(\ref{colormatr}). As the minimally coupled fermions are known to remove the anomaly for Grassmannians~\cite{Abdalla2}, it is reasonable to expect that it also vanishes in the flag manifold case.
\end{itemize}

\noindent
Additional more complicated couplings of the fermions, which would still preserve integrability, are likely possible as well. For example, one could add fermions symmetrically, i.e. by adding to the bosonic chiral Gross-Neveu model an analogous fermionic one, the two parts interacting via a common gauge field. In the undeformed K\"ahler cases (Grassmannians) this is closely related to the supersymmetric version of the model~\cite{Abdalla1}. General quartic fermionic interactions have been investigated in~\cite{Abdalla3}. The $\CP^3$-model with fermions that arises in the context of the $AdS_4\times \CP^3$ superstring was considered in~\cite{BykovAdS, BassoRej}.

\section{{\!A sum rule for flag manifolds $\bullet$ Generalized Einstein metrics}}\label{sumrulesec}

We expect that, in the case of flag manifolds, the Ricci flow equations~(\ref{Ricflowproj}) are also satisfied. The precise determination of all of the ingredients entering those equations is left for the future, and for the moment we will just assume that these equations hold true and derive a rather strong consequence, which may be readily tested. Indeed, let us consider the $\tau\to 0$ ($s\to 1$)  limit of the first equation in~(\ref{Ricflowproj}). First of all, at the homogeneous point the one-form $\mathcal{D} \Phi$ vanishes, as otherwise it would be an $SU(n)$-invariant one-form on a flag manifold, and there are no such one-forms. Near the homogeneous point the metric collapses: $g\simeq (1-s)\,g_{\mathrm{hom.}}$, where $g_{\mathrm{hom.}}$ is a homogeneous (`Killing', reductive) metric on the flag manifold. Since the Ricci tensor is invariant under rescalings of the metric, it converges to the one of the homogeneous metric~$g_{\mathrm{hom.}}$. As the $H$-field is the exterior derivative of the fundamental Hermitian form of the metric, it is proportional to $(1-s)$ as well, and its square $H_{imn}H_{jm'n'}g^{mm'}g^{nn'}$ converges to the value for $g=g_{\mathrm{hom.}}$. Recalling $s=e^{n\tau}$, we have $-\dot{g}\to n\,g_{\mathrm{hom.}}$, so that in the limit the first equation in~(\ref{Ricflowproj}) converges to
\begin{empheq}[box=\fbox]{align}
\hspace{1em}\vspace{1em}
\label{genEinstein}
R_{ij}+{1\over 4}H_{imn}H_{jm'n'}g_{\mathrm{hom.}}^{mm'}g_{\mathrm{hom.}}^{nn'}=n \,(g_{\mathrm{hom.}})_{ij}\,,
\end{empheq}
which is the generalized Einstein condition for the homogeneous metric (in the l.h.s. all quantities are calculated w.r.t. the homogeneous metric). We note that Einstein metrics on homogeneous spaces have been investigated, for instance, in~\cite{Ziller, Alek0, Arvanito}. In particular, the latter paper deals with flag manifolds (we refer the reader to~\cite{Alek} for a detailed discussion of the homogeneous geometry of flag manifolds).

\vspace{0.3cm}\noindent
For symmetric spaces, i.e. Grassmannians $G(m, n)$, the Killing metric is also K\"ahler, so that $H=0$, in which case the above equation is the statement that it is Einstein with cosmological constant $n$. In fact, this is the only case where the Killing metrics are K\"ahler, and the cosmological constant has an interpretation as the degree of the anticanonical bundle. Indeed, recall that for a K\"ahler manifold $\mathcal{M}$ the Ricci form represents the first Chern class $[{i\over 2\pi}\,R_{k\bar{m}}\,dz^k\wedge d\bar{z^m}]=c_1(\mathcal{M})$. In the case of a Grassmannian $G(m, n)$ let us fix the normalization of the metric by the requirement $\mathrm{vol}(\mathcal{C})=1$, where $\mathcal{C}\simeq \CP^1$ is a generator of $H_2(G(m, n), \mathbb{Z})$. In this case the Einstein condition $\mathrm{Ric}=\Lambda \,g$ implies $\int\limits_{\mathcal{C}}\,c_1(\mathcal{M})=\Lambda$. It is known that for Grassmannians $c_1(G(m, n))=n\,\mathcal{C}^\vee$ ($\int\limits_{\mathcal{C}}\,\mathcal{C}^\vee=1$), so that $\Lambda=n$. The most important fact about this result is that the cosmological constant is independent of $m$, i.e. of the denominator of the quotient $U(n)\over U(m)\times U(n-m)$\footnote{From the point of view of $\sigma$-models this may also be seen in a direct calculation of the one-loop $\beta$-function, see~\cite{Zarembo} for a review.}. As equation~(\ref{genEinstein}) shows, a generalization of this statement for arbitrary flag manifolds requires taking into account the generally non-vanishing $H$-field.

\vspace{0.3cm}\noindent
We can now prove~(\ref{genEinstein}) for flag manifolds~(\ref{flaguni}). To this end, we introduce the Maurer-Cartan forms $-g^{-1}dg=\sum\limits_{j, k=1}^s\,E_{jk},$ where $E_{jk}$ for $j\neq k$ are the vielbein one-forms with values in $\mathrm{Hom}(\CC^{n_k}, \CC^{n_j})$. The general invariant metric has the form
\bea
ds^2=\sum\limits_{j\neq k}\,a_{jk}\,\mathrm{Tr}(E_{jk} E_{kj})\,,\quad\quad a_{jk}>0\,.
\eea
The Ricci tensor may as well be expanded in the vielbein: $\scalemath{0.8}{ \sum\limits_{\mu, \nu}\,R_{\mu\nu}dx^\mu dx^\nu= \sum\limits_{j\neq k}\,R_{jk}\,\mathrm{Tr}(E_{jk} E_{kj})}$. An expression for $R_{jk}$ is found in~\cite{Arvanito}. In our case $a_{jk}=1$, which gives 
\bea\label{Rij}
R_{jk}=\frac{n+n_j+n_k}{2}\,,\quad\quad 1 \leq j, k \leq s\,.
\eea
The metric is Einstein if $R_{jk}=\Lambda=\mathrm{const.}$ There are two cases when this happens:
\begin{itemize}
\item Grassmannians, i.e. $s=2$, $n_1+n_2=n$, $R_{jk}=n$
\item Flag manifolds, for which $n_j=p=\mathrm{const.}$, $n=ps$. In this case $R_{jk}=\frac{n+2p}{2}$. For complete flags $p=1$ and $R_{jk}=\frac{n+2}{2}$.
\end{itemize}
We come to the computation of $H$. The flatness of $g^{-1}dg$ gives $DE_{jk}=\sum\limits_{p\neq j, k}\,E_{jp}\wedge E_{pk}$, where $D$ is the $\prod\limits_{j=1}^s U(n_j)$-covariant derivative. Since the $B$-field is $\scalemath{0.8}{ B=i\,\sum\limits_{j\neq k}\,{b_{jk}\over 2}\,\mathrm{Tr}(E_{jk}\wedge E_{kj})}$ with $b_{jk}=\pm1 \,(j\lessgtr k)$, we find 
\bea
H=dB=i\,\sum\limits_{j\neq k\neq m}\,b_{jk}\,\mathrm{Tr}(E_{jm}\wedge E_{mk}\wedge E_{kj})\,,
\eea

\vspace{-0.3cm}\noindent
so that, in double-index notations, the nonzero components are $H_{jm|mk|kj}=b_{jk}+b_{km}+b_{mj}=\pm 1$, and permutations thereof. The metric has the form $\langle (E_{jm})_{a\alpha}, (E_{nk})_{\beta b}\rangle=\delta_{jk}\delta_{mn}\delta_{ab}\delta_{\alpha\beta}$ ($a, b=1\ldots n_j$, $\alpha, \beta = 1\ldots n_m$), so that
\bea
(H^2)_{jk|mn}=2\,\delta_{jn}\delta_{km}\,\sum\limits_{p\neq j, k}\,n_p\,H_{jk|kp|pj}\,H_{kj|pk|jp}=2\,\delta_{jn}\delta_{km}\,(n-n_j-n_k)\,.
\eea

\vspace{-0.3cm}\noindent
Uniformizing notation with~(\ref{Rij}), we define $(H^2)_{jk|mn}:=\delta_{jn}\delta_{km}\,(H^2)_{jk}$, where $(H^2)_{jk}=2\,(n-n_j-n_k)$. As a result, we obtain the \emph{sum rule}
\bea
R_{jk}+{1\over 4}\,(H^2)_{jk}=n\,,
\eea
which is equivalent to the generalized Einstein condition~(\ref{genEinstein}). The key point is that the sum depends only on the dimension of the ambient space $\CC^n$, but not on the structure of the flags~(\ref{flaguni}).

\section{Conclusion and outlook}

In the present paper we studied $\sigma$-models with complex homogeneous target spaces (projective space $\CP^{n-1}$ and, more generally, flag manifolds), as well as their trigonometric deformations. The deformations in question are rather special in that they preserve the Hermiticity of the metric, at the expense of introducing additional fields, such as a two-form field $B$ and the dilaton $\Phi$. It was shown that the models may in fact be formulated as gauged bosonic Thirring-like systems. We started by elaborating the ungauged version in section~\ref{betagammaUG}, computing the one-loop $\beta$-function and showing that the Ricci flow equation is satisfied by a simple evolution of the deformation parameter. In fact, the $\beta$-function~(\ref{betafunc}), (\ref{betafunc2}) turns out to be universal for gauged and ungauged models alike. This has rather strong consequences, one of them being that the $\beta$-functions of all $\sigma$-models of flags in $\CC^n$ depend only on $n$, and not on the structure of the flag. We have termed this `a sum rule for flag manifolds'. In the case of Grassmannians $G(m, n)$ this is the familiar statement that the one-loop $\beta$-function is independent of $m$. We have also checked that the Ricci flow equations are satisfied for the $\CP^{n-1}$ models\footnote{The verification was done numerically for low values of $n$.}, when one adds an additional compensating gauge field $\mathcal{E}$, whose meaning has also been explained. 

\vspace{0.3cm}\noindent
The crucial difference between the gauged and ungauged versions of the model (which correspond to $\CP^{n-1}$ and $\CC^n$ models respectively) is that the symmetry that one wishes to gauge is in fact chiral, and it is violated by quantum anomalies. To get rid of those, one may couple the bosonic Thirring-like system to fermions, as in~(\ref{cpnferm}) or~(\ref{flaglagr}). Incidentally these are the same fermionic couplings that cancel the anomaly in the so-called L\"uscher non-local charge~\cite{Luscher}. This anomaly is otherwise present in the $\CP^{n-1}$ and $G(m, n)$ models~\cite{Abdalla1, Abdalla2}. This leads us to conjecture that the flag manifold $\sigma$-models with fermions are quantum integrable models.

\vspace{0.3cm}\noindent
The models of the type considered in the present paper possess yet another curious property. Apparently their Lax connections have ultralocal Poisson brackets: this was observed in the context of Hermitian symmetric spaces in~\cite{Bytsko,Zagermann}\footnote{In the case of Hermitian symmetric spaces, and in the homogeneous/rational situation, the $B$-field is topological, but the Lax pair constructed according to Pohlmeyer's procedure is different from the conventional Lax pair in the absence of the $B$-field~\cite{BykovZeroCurv}.} and shown generally for a related class of para-Hermitian geometries in~\cite{DelducZT}\footnote{Ultralocal Lax pairs for the deformed $\CP^1$-model were constructed in~\cite{Lukyanov, Kotousov}.}. It was observed in~\cite{Zagermann} that such peculiar properties of models with Hermitian symmetric space geometries may be attributed to their origins in the Ashtekar variables of general relativity, upon dimensional reduction. Another parallel with the Ashtekar variables is that the $\sigma$-models in question have only polynomial interactions in the Thirring-like formulation~(cf.~(\ref{inhomgaugelagr})). We leave the further elucidation of these ideas for future work.

\vspace{0.3cm}\noindent
\textbf{Acknowledgments.} I would like to thank A.~A.~Slavnov for support and G.~Arutyunov, D.~L\"ust for discussions. I am especially grateful to K.~Zarembo for reading the manuscript and many useful remarks and suggestions.

\vspace{1.5cm}\noindent
\whiterem{ \vspace{0.3cm}\quad {\large Appendix}\quad}
\addappheadtotoc
\begin{appendices}

\vspace{-0.4cm}\noindent
\section{Metric, $B$-field and dilaton of the deformed $\CC^n$-model}\label{appRicci}

Here we derive the formulas~(\ref{Cnmetricdef1}) from the main text. We start by writing out the quadratic form in the Lagrangian~(\ref{lagr11}):
\bear\label{Vquadform}
&&\mathrm{Tr}\left(r_s(UV)(UV)^\dagger\right)=\sum\limits_{k, m=1}^n\,V_k \bar{V}_m \underbracket[0.6pt][0.6ex]{\left(\frac{\delta_{km}}{\hat{\lambda}_k}+(b-{\upgamma\over n})\,U_k \bar{U}_m\right)}_{:=Q_{km}}\,,\\ \nonumber
&&\textrm{where}\quad\quad \hat{\lambda}_m^{-1}=\upalpha\,\sum_{j<m}\,|U_j|^2+\upbeta\,\sum_{j>m}\,|U_j|^2+\upgamma\,|U_m|^2\,.
\eear
The constants $\upalpha, \upbeta, \upgamma$ were defined in~(\ref{acoefs}). Inverting the quadratic form $Q$, we obtain the Lagrangian of the $\sigma$-model:
\bea
\mathrsfso{L}=\sum\limits_{k, m=1}^n\,\dd \bar{U}_k \left(\delta_{km}{\hat{\lambda}_k}-\frac{\hat{\lambda}_k \hat{\lambda}_m\,U_k \bar{U}_m}{\upkappa+\sum\limits_p\,\hat{\lambda}_p\,|U_p|^2}\right) \bd U_m,\quad\quad \upkappa=(b-{\upgamma\over n})^{-1}\,.
\eea
It is convenient to introduce the parametrization $U_j=e^{t_j- i\,\phi_j}$ and the notation $\lambda_j:=\hat{\lambda}_j \,e^{2t_j}$ (we already used it in sec.~\ref{geomapprsec}). In this case $d U_j\wedge d\bar{U}_j=2i\,e^{2t_j}\,dt_j\wedge d\phi_j$ and $\sum\limits_{j=1}^n\,\hat{\lambda}_j\,U_j \,d\bar{U}_j=\sum\limits_{j=1}^n\,\lambda_j\,(dt_j- i\,d\phi_j)$. Therefore\footnote{We introduce a factor of $2$ in the definition of $k_{ij}$ so that the metric satisfies the canonically normalized Ricci flow equations~(\ref{Ricflowaff}). One other way to fix this normalization is via the Einstein condition~(\ref{genEinstein}) for undeformed metrics.}
\bear\label{affinemetr}
&&ds^2=
\sum\limits_{i, j=1}^n\,k_{ij} (dt_i\,dt_j+d\phi_i d\phi_j)\,.\\ \label{affineBfield}
&&B= \sum\limits_{i, j=1}^n\,k_{ij}\, dt_i\wedge d\phi_j\,,\quad\quad
\textrm{where}\quad\quad {1\over 2}\,k_{ij}=\lambda_i\,\delta_{ij}-\frac{\lambda_i \lambda_j}{\upkappa+\sum\limits_k \lambda_k}\,.
\eear
The dilaton $\Phi$ arises from the integration over the $V$-variables, namely from the determinant of the quadratic form in~(\ref{Vquadform}):
\bea\label{dilaffine}
e^{-2\Phi}=\mathrm{Det}(Q)=\prod\limits_{k=1}^n\,{1\over \hat{\lambda}_k}\times \left(1+{1\over \upkappa}\,\sum\limits_{k=1}^n\hat{\lambda}_k |U_k|^2\right)\,.
\eea

\subsection[$n=2$: the deformed $\CC^2$]{$n=2$: the deformed $\CC^2$.}
In computing the Ricci tensor and other geometric quantities it is much more convenient to deal with the inverse metric $\kappa^{-1}$, which has the simple form
\bea\label{kappainv}
\kappa^{ij}={1\over 2}\underbracket[0.6pt][0.6ex]{\lambda_i^{-1}}_{:=\upchi_i}\,\delta^{ij}+{1\over 2\upkappa}\,.
\eea
After the shift of $t_i$-variables described in sec.~\ref{betafuncsec} the functions $\upchi_i$ are found to be
\bea\label{chifunc}
\upchi_1={1\over 2}\,{1+e^{2\tau}\over 1-e^{2\tau}}+{e^{\tau}\over 1-e^{2\tau}}\,e^{t_2-t_1},\quad\quad \upchi_2={1\over 2}\,{1+e^{2\tau}\over 1-e^{2\tau}}+{e^{\tau}\over 1-e^{2\tau}}\,e^{t_1-t_2}\,.
\eea
We start by computing the angular components $R^{\phi_i \phi_j}$ (again, with upper indices) of the Ricci tensor. These may be obtained using the formula $R^\mu_{\nu}\,K^\nu={1\over \sqrt{g}}\,\dd_\rho \left(\sqrt{g}\,\nabla^{\mu}K^\rho\right)$, valid for any Killing vector field $K$ (cf.~\cite{Blau}). As it follows from~(\ref{kappainv}), ${\dd \kappa^{ij}\over \dd t_k}={1\over 2}{\dd \upchi_i\over \dd t_k}\,\delta^{ij}$. Using this as well as the fact that $\upchi_i=\upchi_i(t_1-t_2)$, we arrive at the formula for the $\phi$-components of the Ricci tensor:
\bea
R^{\phi_i \phi_j}=\sum\limits_{p=1}^n\,{1\over 8k}\,{\dd \over \dd t_p}\left(k\,\upchi_p\,{\dd \upchi_i\over \dd t_p}\right)\,\delta_{ij}-k_{ij}\,\sum\limits_{p=1}^n\,{1\over 8}\,\upchi_p\,{\dd \upchi_i\over \dd t_p} \,{\dd \upchi_j\over \dd t_p}\,,\quad\quad k=\mathrm{det}(k_{ij})\,.
\eea
For other ingredients of the r.h.s. of~(\ref{Ricflowaff}) we also obtain expressions in terms of $\upchi_i$:
\bear
(H^2)^{\phi_i \phi_j}=-{1\over 2}{\dd \upchi_i\over \dd t_j}\,{\dd \upchi_j\over \dd t_i}+{1\over 2}\,k_{ij}\,\sum\limits_{p=1}^n\,\upchi_p\,{\dd \upchi_i\over \dd t_p} \,{\dd \upchi_j\over \dd t_p}\,\\
2\,\nabla^{\phi_i} \nabla^{\phi_j} \Phi={1\over 8}\,\delta^{ij}\,\sum\limits_{p=1}^n\,\upchi_p\,{\dd \upchi_i\over \dd t_p}\,\left(2-{1\over k}\,{\dd k\over \dd t_p}\right)\,.
\eear
Here we have used that the dilaton is $\Phi=-t_1-t_2+{1\over 2}\log{k}$. Assembling the pieces, we arrive at the following expression for the r.h.s of the Ricci flow equation~(\ref{Ricflowaff}):
\bea
\scalemath{0.88}{
R^{\phi_i \phi_j}+{1\over 4}\,(H^2)^{\phi_i \phi_j}+2\,\nabla^{\phi_i} \nabla^{\phi_j} \Phi=\sum\limits_{p=1}^n\,\left({1\over 8}\,{\dd \over \dd t_p}\left(\,\upchi_p\,{\dd \upchi_i\over \dd t_p}\right)+{1\over 4}\upchi_p\,{\dd \upchi_i\over \dd t_p}\right)\,\delta_{ij}-{1\over 8}\,{\dd \upchi_i\over \dd t_j}\,{\dd \upchi_j\over \dd t_i}\,.}
\eea
Using the explicit expressions~(\ref{chifunc}), we find that
\bea
{d \kappa^{ij}\over d\tau}=R^{\phi_i \phi_j}+{1\over 4}\,(H^2)^{\phi_i \phi_j}+2\,\nabla^{\phi_i} \nabla^{\phi_j} \Phi\,.
\eea

\noindent
The mixed $(t, \phi)$-components of both sides of the Ricci flow equations are identically zero, so it remains to compute the $(t, t)$-components. Here we do not find any significant simplifications, so we just outline the strategy of how this can be done. The $(t, t)$-components of the Ricci tensor may be written as follows:
\bea
R_{t_i t_j}=R_{ij}^{(k)}-{1\over 2}\,\nabla_i \nabla_j \,\log{k}-{1\over 4}\,\mathrm{Tr}\left({\dd k^{-1}\over \dd t_i}\,k\, {\dd k^{-1}\over \dd t_j}\,k\right)\,
\eea
where $R_{ij}^{(k)}$ is the Ricci tensor of the two-dimensional metric $k_{ij}$. As for any such metric, $R_{ij}^{(k)}={R\over 2}\,k_{ij}$, where $R$ is the Ricci scalar. The latter can be computed by noticing that the metric has additional isometry $t_i\to t_i+\mathrm{const.}$ Introducing the variables $x=t_1-t_2$ and $y=t_1+t_2$ and shifting $y$ to remove the cross-term, we bring the metric to diagonal form:
\bea
k_{ij}\,dt_i\,dt_j=\frac{2\,dx^2}{\upchi_1+\upchi_2}+{k\over 2}\,(\upchi_1+\upchi_2)\,dy^2\,,\quad\quad k(x)={\upkappa\over \upchi_1+\upchi_2+\upkappa\,\upchi_1\upchi_2}\,.
\eea
Denoting $\upchi_1+\upchi_2:=c(x)$, we calculate the Ricci scalar $2R=-c''-{k''\over k}\,c-{3\over 2}\,{k'\over k}\,c'+{c\over 2}\,{(k')^2\over k^2}$. One also has a simple formula for the Christoffel symbols with all upper indices: $\Gamma^{p|mn}=-{1\over 8} \upchi_n\,\dd_n\upchi_m\,\delta^{pm}-{1\over 8} \upchi_m\,\dd_m\upchi_n\,\delta^{pn}+{1\over 8}\,\upchi_p\,\dd_p\upchi_n\,\delta^{mn}$. The rest is a matter of direct calculation, which shows that the Ricci flow equation~(\ref{Ricflowaff}) is satisfied. It would be interesting to find a proper geometrical framework that would facilitate such calculations.

\section{Computing the $\beta$-function in the inhomogeneous gauge}\label{betafuncinhomspp}

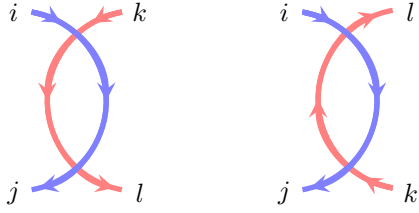
\begin{figure}
\centering
\bea\nonumber
\begin{tikzpicture}[
baseline=-\the\dimexpr\fontdimen22\textfont2\relax,scale=1.2]
\draw [-stealth, red!50, line width=2pt] (2,0) arc [radius=1, start angle=100, end angle= 120];
\draw [-stealth, red!50, line width=2pt] (2,0) arc [radius=1, start angle=100, end angle= 180];
\draw [-stealth, red!50, line width=2pt] (2,0) arc [radius=1, start angle=100, end angle= 255];
\draw [red!50, line width=2pt] (2,0) arc [radius=1, start angle=100, end angle= 260];
\draw [-stealth, blue!50, line width=2pt] (1,0) arc [radius=1, start angle=80, end angle= 60];
\draw [-stealth, blue!50, line width=2pt] (1,0) arc [radius=1, start angle=80, end angle= 0];
\draw [-stealth, blue!50, line width=2pt] (1,0) arc [radius=1, start angle=80, end angle= -75];
\draw [blue!50, line width=2pt] (1,0) arc [radius=1, start angle=80, end angle= -80];
\node at (2.2,0) {\footnotesize $k$};
\node at (0.8,0) {\footnotesize $i$};
\node at (0.8,-2) {\footnotesize $j$};
\node at (2.2,-2) {\footnotesize $l$};
\draw [-stealth, red!50, line width=2pt] (5,-1.95) arc [radius=1, start angle=260, end angle= 240];
\draw [-stealth, red!50, line width=2pt] (5,-1.95) arc [radius=1, start angle=260, end angle= 180];
\draw [-stealth, red!50, line width=2pt] (5,-1.95) arc [radius=1, start angle=260, end angle= 110];
\draw [red!50, line width=2pt] (5,-1.95) arc [radius=1, start angle=260, end angle= 100];
\draw [-stealth, blue!50, line width=2pt] (4,0) arc [radius=1, start angle=80, end angle= 60];
\draw [-stealth, blue!50, line width=2pt] (4,0) arc [radius=1, start angle=80, end angle= 0];
\draw [-stealth, blue!50, line width=2pt] (4,0) arc [radius=1, start angle=80, end angle= -75];
\draw [blue!50, line width=2pt] (4,0) arc [radius=1, start angle=80, end angle= -80];
\node at (5.2,0) {\footnotesize $l$};
\node at (3.8,0) {\footnotesize $i$};
\node at (3.8,-2) {\footnotesize $j$};
\node at (5.2,-2) {\footnotesize $k$};
\end{tikzpicture}
\eea
\caption{Diagrams contributing to $\left(\beta_{ij}^{kl}\right)_{1}$. Here $i, j, k, l=1, \ldots, n-1$.} \label{fig4}
\end{figure}

In this appendix we consider the renormalization of the theory~(\ref{inhomgaugelagr}) at one loop. We wish to show that all divergences may be reabsorbed in a redefinition of the deformation parameter $s$ entering $\upalpha, \upbeta, \upgamma$. The quartic vertices in~(\ref{inhomgaugelagr}) look exactly as in~(\ref{lagr6}), but with the summation restricted to $1 \ldots n-1$ and the truncated $r$-matrix of the form
\bea
\tilde{r}_{ij}^{kl}=a_{ij}\delta_i^k \delta_j^l+\upgamma\,\delta_{ij}\delta^{kl},\quad\quad i, j, k, l=1\ldots n-1\,.
\eea
One contribution to the $\beta$-function comes from exactly the same diagrams as before, shown in~Fig.~\ref{fig4}, where in the vertices one replaces $r\to \tilde{r}$:
\bear
&&\left(\beta_{ij}^{kl}\right)_{1}=\left[\frac{(n-1)s}{(1-s)^2}+(i-j)a_{ij}\right]\,\left(\delta_i^k \delta_j^l-{1\over n-1}\delta_{ij}\delta^{kl}\right)=\\ \nonumber
&&=\left[(n-1)\,\upalpha\upbeta+(i-j)a_{ij}\right]\,\delta_i^k \delta_j^l-\upalpha\upbeta\,\delta_{ij}\delta^{kl}\,.
\eear

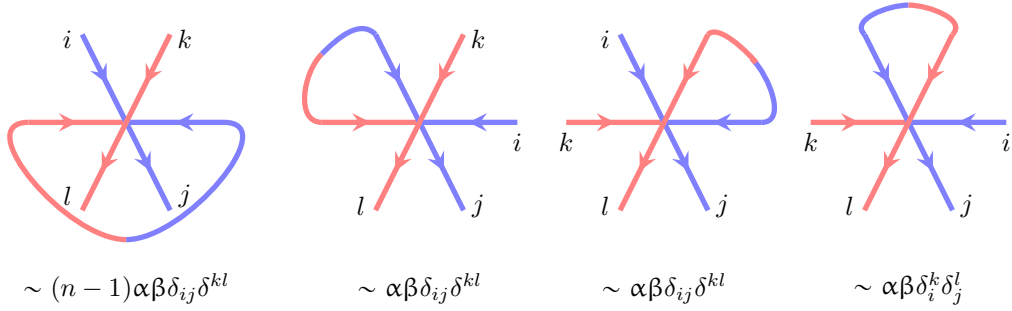
\begin{figure}
\centering
\bea \nonumber
\begin{tikzpicture}[
baseline=-\the\dimexpr\fontdimen22\textfont2\relax,scale=1.3]
\draw[-stealth, red!50, line width=2pt, rounded corners] (-1,0)  --  (-0.5,0) node[right,black] {};
\draw[red!50, line width=2pt, rounded corners] (-0.6,0)  --  (0,0) node[right,black] {};
\draw[blue!50, line width=2pt, rounded corners] (0,0)  --  (0.6,0) node[right,black] {};
\draw[-stealth, blue!50, line width=2pt, rounded corners] (1,0)  -- (0.5,0)   node[right,black] {};
\draw[-stealth, blue!50, line width=2pt, rounded corners] (-0.45,0.9)  --  (-0.2,0.4) node[right,black] {};
\draw[-stealth, blue!50, line width=2pt, rounded corners] (-0.25,0.5) -- (0.25,-0.5)    node[right,black] {};
\draw[blue!50, line width=2pt, rounded corners] (0.2,-0.4) -- (0.45,-0.9)    node[right,black] {};

\draw[-stealth, red!50, line width=2pt, rounded corners] (0.45,0.9)  --  (0.2,0.4) node[right,black] {};
\draw[-stealth, red!50, line width=2pt, rounded corners] (0.25,0.5) -- (-0.25,-0.5)    node[right,black] {};
\draw[red!50, line width=2pt, rounded corners] (-0.2,-0.4) -- (-0.45,-0.9)    node[right,black] {};

\node at (-0.6,0.85) {\footnotesize $i$};
\node at (0.6,0.85) {\footnotesize $k$};
\node at (0.6,-0.75) {\footnotesize $j$};
\node at (-0.6,-0.75) {\footnotesize $l$};

\draw[red!50, line width=2pt, rounded corners] (-1,0) to [out=180,in=180] (0,-1.2);
\draw[blue!50, line width=2pt, rounded corners] (0,-1.2) to [out=0,in=0] (1,0);

\node at (0,-1.7) {\footnotesize $\sim (n-1)\upalpha\upbeta \delta_{ij}\delta^{kl}$};


\draw[-stealth, red!50, line width=2pt, rounded corners] (2,0)  --  (2.5,0) node[right,black] {};
\draw[red!50, line width=2pt, rounded corners] (2.4,0)  --  (3,0) node[right,black] {};
\draw[blue!50, line width=2pt, rounded corners] (3,0)  --  (3.6,0) node[right,black] {};
\draw[-stealth, blue!50, line width=2pt, rounded corners] (4,0)  -- (3.5,0)   node[right,black] {};
\draw[-stealth, blue!50, line width=2pt, rounded corners] (2.55,0.9)  --  (2.8,0.4) node[right,black] {};
\draw[-stealth, blue!50, line width=2pt, rounded corners] (2.75,0.5) -- (3.25,-0.5)    node[right,black] {};
\draw[blue!50, line width=2pt, rounded corners] (3.2,-0.4) -- (3.45,-0.9)    node[right,black] {};

\draw[-stealth, red!50, line width=2pt, rounded corners] (3.45,0.9)  --  (3.2,0.4) node[right,black] {};
\draw[-stealth, red!50, line width=2pt, rounded corners] (3.25,0.5) -- (2.75,-0.5)    node[right,black] {};
\draw[red!50, line width=2pt, rounded corners] (2.8,-0.4) -- (2.55,-0.9)    node[right,black] {};

\node at (4,-0.2) {\footnotesize $i$};
\node at (3.6,0.85) {\footnotesize $k$};
\node at (3.6,-0.85) {\footnotesize $j$};
\node at (2.4,-0.85) {\footnotesize $l$};

\draw[red!50, line width=2pt, rounded corners] (2,0) to [out=180,in=-135] (2,0.7);
\draw[blue!50, line width=2pt, rounded corners] (2.55,0.9) to [out=135,in=45] (2,0.7);

\node at (3,-1.7) {\footnotesize $\sim \upalpha\upbeta \delta_{ij}\delta^{kl}$};


\draw[-stealth, red!50, line width=2pt, rounded corners] (4.5,0)  --  (5,0) node[right,black] {};
\draw[red!50, line width=2pt, rounded corners] (4.9,0)  --  (5.5,0) node[right,black] {};
\draw[blue!50, line width=2pt, rounded corners] (5.5,0)  --  (6.1,0) node[right,black] {};
\draw[-stealth, blue!50, line width=2pt, rounded corners] (6.5,0)  -- (6,0)   node[right,black] {};
\draw[-stealth, blue!50, line width=2pt, rounded corners] (5.05,0.9)  --  (5.3,0.4) node[right,black] {};
\draw[-stealth, blue!50, line width=2pt, rounded corners] (5.25,0.5) -- (5.75,-0.5)    node[right,black] {};
\draw[blue!50, line width=2pt, rounded corners] (5.7,-0.4) -- (5.95,-0.9)    node[right,black] {};

\draw[-stealth, red!50, line width=2pt, rounded corners] (5.95,0.9)  --  (5.7,0.4) node[right,black] {};
\draw[-stealth, red!50, line width=2pt, rounded corners] (5.75,0.5) -- (5.25,-0.5)    node[right,black] {};
\draw[red!50, line width=2pt, rounded corners] (5.3,-0.4) -- (5.05,-0.9)    node[right,black] {};

\node at (4.5,-0.2) {\footnotesize $k$};
\node at (4.9,0.85) {\footnotesize $i$};
\node at (6.1,-0.85) {\footnotesize $j$};
\node at (4.9,-0.85) {\footnotesize $l$};

\draw[blue!50, line width=2pt, rounded corners] (6.5,0) to [out=0,in=-45] (6.4,0.65);
\draw[red!50, line width=2pt, rounded corners] (5.95,0.9) to [out=45,in=-45] (6.4,0.65);

\node at (5.5,-1.7) {\footnotesize $\sim \upalpha\upbeta \delta_{ij}\delta^{kl}$};


\draw[-stealth, red!50, line width=2pt, rounded corners] (7,0)  --  (7.5,0) node[right,black] {};
\draw[red!50, line width=2pt, rounded corners] (7.4,0)  --  (8,0) node[right,black] {};
\draw[blue!50, line width=2pt, rounded corners] (8,0)  --  (8.6,0) node[right,black] {};
\draw[-stealth, blue!50, line width=2pt, rounded corners] (9,0)  -- (8.5,0)   node[right,black] {};
\draw[-stealth, blue!50, line width=2pt, rounded corners] (7.55,0.9)  --  (7.8,0.4) node[right,black] {};
\draw[-stealth, blue!50, line width=2pt, rounded corners] (7.75,0.5) -- (8.25,-0.5)    node[right,black] {};
\draw[blue!50, line width=2pt, rounded corners] (8.2,-0.4) -- (8.45,-0.9)    node[right,black] {};

\draw[-stealth, red!50, line width=2pt, rounded corners] (8.45,0.9)  --  (8.2,0.4) node[right,black] {};
\draw[-stealth, red!50, line width=2pt, rounded corners] (8.25,0.5) -- (7.75,-0.5)    node[right,black] {};
\draw[red!50, line width=2pt, rounded corners] (7.8,-0.4) -- (7.55,-0.9)    node[right,black] {};

\node at (7,-0.2) {\footnotesize $k$};
\node at (9,-0.2) {\footnotesize $i$};
\node at (8.6,-0.85) {\footnotesize $j$};
\node at (7.4,-0.85) {\footnotesize $l$};

\draw[blue!50, line width=2pt, rounded corners] (7.55,0.9) to [out=135,in=180] (8,1.2);
\draw[red!50, line width=2pt, rounded corners] (8.45,0.9) to [out=45,in=0] (8,1.2);

\node at (8,-1.7) {\footnotesize $\sim \upalpha\upbeta \delta_i^k\delta_j^{l}$};

\end{tikzpicture}
\eea
\vspace{-0.5cm}
\caption{Bubbles in the sextic vertex, contributing to $\left(\beta_{ij}^{kl}\right)_{2}$.} \label{fig5}
\end{figure}
\vspace{-0.7cm}\noindent
An additional contribution comes from bubbles in the sextic vertex (see Fig.~\ref{fig5}):
\bea
\left(\beta_{ij}^{kl}\right)_{2}=\upalpha\upbeta(n+1)\,\delta_{ij}\delta^{kl}+\upalpha\upbeta\delta_i^k \delta_j^l\,.
\eea
Summing the two contributions, we obtain
\bea
\beta_{ij}^{kl}=\left[n\,\upalpha\upbeta+(i-j)a_{ij}\right]\,\delta_i^k \delta_j^l+n\upalpha\upbeta\,\delta_{ij}\delta^{kl}
\eea
As explained in sec.~\ref{betafuncsec}, the part proportional to $(i-j)a_{ij}$ can be removed by a redefinition of coordinates, and the remaining terms give rise to the Ricci flow equation $\dot{\upgamma}=n\upalpha\upbeta$ that is solved by $s=e^{n\tau}$.

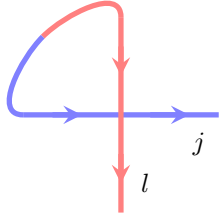
\begin{figure}
\centering
\bea\nonumber
\begin{tikzpicture}[
baseline=-\the\dimexpr\fontdimen22\textfont2\relax,scale=1.3]
\draw[-stealth, blue!50, line width=2pt, rounded corners] (5,0)  -- (5.6,0) node[right,black] {};
\draw[-stealth, blue!50, line width=2pt, rounded corners] (5.3,0)  -- (6.7,0) node[right,black] {};
\draw[blue!50, line width=2pt, rounded corners] (6.3,0)  -- (7,0) node[right,black] {};
\draw[-stealth, red!50, line width=2pt, rounded corners] (6,1)  -- (6,0.4) node[right,black] {};
\draw[-stealth, red!50, line width=2pt, rounded corners] (6,0.7)  -- (6,-0.7) node[right,black] {};
\draw[red!50, line width=2pt, rounded corners] (6,-0.4)  -- (6,-1) node[right,black] {};
\node at (6.8,-0.3) {\footnotesize $j$};
\node at (6.25,-0.7) {\footnotesize $l$};

\draw[red!50, line width=2pt, rounded corners] (6,1) to [out=90,in=45] (5.2,0.8);
\draw[blue!50, line width=2pt, rounded corners] (5,0) to [out=180,in=-135] (5.2,0.8);

\end{tikzpicture}
\eea
\vspace{-0.7cm}
\caption{A bubble in the quartic vertex, contributing to $\upbeta \,|V_k|^2$ terms.} \label{fig6}
\end{figure}

\vspace{0.3cm}\noindent
As another illustration let us consider the renormalization of the term $\upbeta \,|V_k|^2$ in~(\ref{inhomgaugelagr}). The divergent graphs that contribute to the $\beta$-function of this `vertex' come from bubbles in the quartic vertex, see Fig.~\ref{fig6}:
\bea\label{betabeta}
\beta_{i}^j=\delta_{i}^j\,\upbeta\,\sum\limits_{k=1}^{n-1}\,\tilde{r}_{\,ki}^{\,kj}=\delta_{i}^j\,\upbeta\,\left(\sum\limits_{k=1}^{n-1}\,a_{kj}+\upgamma\right)=\delta_{i}^j\,\upbeta\,((n-j)+\upalpha\, n)
\eea
The first term in brackets is again cancelled by a redefinition of variables. We showed in sec.~\ref{betafuncsec} that the $U$-variables evolve with the Ricci flow as $|U_j|^2=e^{j\,\tau}$. However, since we have chosen a gauge $U_n=1$, we have to perform a compensating gauge transformation, and as a result the evolution of the variables is given by $|U_j|^2=e^{(j-n)\,\tau}$. The $V$-variables evolve as inverses of these, $|V_j|^2=e^{-(j-n)\,\tau}$, and this eliminates the unwanted term in~(\ref{betabeta}). The remaining flow equation is $\dot{\upbeta}=n\upalpha\upbeta$, which is again solved by $s=e^{n\tau}$. A similar analysis could be performed for other vertices in the Lagrangian~(\ref{inhomgaugelagr}).

\section{The deformed flag manifold $\sigma$-model in geometric form}\label{defflagmetrapp}

We start this appendix by showing that there is a GLSM formulation of the flag manifold $\sigma$-model based on $n\times n$-matrices rather than on $m\times n$ matrices as in Sec.~\ref{defflagsec}. The quotient~(\ref{flagcomplex}) suggests the following GLSM representation for a deformed flag manifold $\sigma$-model:
\bea\label{lagrGL}
\mathrsfso{L}=
\mathrm{Tr}\left(V \bar{\mathscr{D}} U\right)-\mathrm{Tr}\left(V \bar{\mathscr{D}} U\right)^\dagger+\mathrm{Tr}\left((r_s^{\mathrm{T}})^{-1}(\Phi) \bar{\Phi}\right)
\eea
Here $U, V, \Phi\in \mathrm{End}(\CC^n)$ and the covariant derivatives are defined as follows: $\bar{\mathscr{D}}U=\bd U+i\,\bar{\Phi} U+i \,U \bar{\mathcal{A}}$, $\bar{D}U=\bd U+i \,U \bar{\mathcal{A}}$.  As we now explain, one can simplify this gauged theory to arrive at the GLSM formulation of~(\ref{lagrGL2}) if one assumes that $U$ is non-degenerate, i.e. $U\in GL(n, \CC)$. Taking the variation of~(\ref{lagrGL}) w.r.t. $\bar{\mathcal{A}}$, we find $VU|_{\mathfrak{p}^\vee}=0$ (here we view $VU\in \mathfrak{g}^\vee$). In general this is a rather non-trivial condition, however already from the requirement of vanishing of the first $n_s$ rows of the matrix $VU$ we find that the first $n_s$ rows of $V$ are orthogonal to all columns of~$U$. Since $U$ is non-generate, it follows that $V_{ik}=0$ for $i=1, \ldots, n_s$. Therefore we can safely truncate the matrix $V$ by erasing its first $n_s$ rows, forming a reduced matrix $\tilde{V}\in \mathrm{Hom}(\CC^n, \CC^m)$ with $m=\sum\limits_{i=1}^{s-1}\,n_i$. It is then easy to see that the first $n_s$ columns of the matrix $U$ do not enter the Lagrangian~(\ref{lagrGL}) either, so we may analogously truncate the matrix $U$, forming a new matrix $\tilde{U}\in \mathrm{Hom}(\CC^m, \CC^n)$. The Lagrangian then takes the form
\bea\label{lagrGL3}
\mathrsfso{L}= \mathrm{Tr}\left(\tilde{V} \bar{\mathscr{D}} \tilde{U}\right)-\mathrm{Tr}\left(\tilde{V} \bar{\mathscr{D}} \tilde{U}\right)^\dagger+\mathrm{Tr}\left((r_s^{\mathrm{T}})^{-1}(\Phi) \bar{\Phi}\right)\,,
\eea
where it is understood that the gauge field in the covariant derivatives has been truncated accordingly. If one drops the tildes, one recognizes formula~(\ref{lagrGL2}).

\vspace{0.3cm}\noindent
Our next goal is to derive the metric form of the deformed flag manifold model~(\ref{lagrGL3}). From now on we will drop the tildes over $U$ and $V$ but we will still assume that they are $m\times n$-matrices. If necessary, one can always return from~(\ref{lagrGL3}) back to~(\ref{lagrGL}) by setting $m=n$, so in a sense the $m\neq n$ notation is more general. Our goal in this section is to eliminate the variables $V, \bar{V}$, which enter the Lagrangian quadratically. The relevant equations to be solved are~(\ref{Veq1})-(\ref{Veq2}) with $\hat{r}=r$. To write down the solution, we introduce the projector $\Pi_U=U(\bar{U} U)^{-1}\bar{U}$ on $\mathrm{Im}(U)$ (indeed $\Pi_U(UY)=UY$). Then
\bea\label{VCsol}
V={1\over \bar{U} U} \,\bar{U} \left[\frac{1}{\frac{1}{2}\,\frac{1+s}{1-s}\,\mathrm{Id}+\Pi_U\,{i\over 2}\,\mathcal{R}}\right]\, \left(U {1\over \bar{U} U} D\bar{U} \right)\,.
\eea
It is easy to prove that in the denominator one can freely commute $\Pi_U$ with $\mathcal{R}$. The formula is proven by noting that $\bar{U}\,r_s\,\Pi_U=\bar{U} \left(\frac{1}{2}\,\frac{1+s}{1-s}\,\mathrm{Id}+{i\over 2}\,\mathcal{R} \,\Pi_U\, \right)$ and using this commutation property.

\vspace{0.3cm}\noindent
The formula~(\ref{VCsol}) for $V$ can be considerably simplified by using a part of the complex gauge symmetry. We will call this procedure `passing to the unitary frame', for the following reason. We may write $U=\hat{U}\hat{b}$, where $\bar{\hat{U}} \hat{U}=\mathds{1}_m$ and $\hat{b}\in B\mysub P_{m_1, \ldots, m_s}$ is a lower-triangular matrix ($B$ is the Borel subgroup). This decomposition is simply the Gram-Schmidt orthogonalization of the $m$ vectors in $U$. The remaining gauge group is then $H=U(n_1)\times \cdots \times U(n_s)$, which is of course the denominator of the quotient~(\ref{flaguni}). The solution for $V$ clearly simplifies in the unitary gauge $\bar{U} U=\mathds{1}_m$. Using this gauge and inserting $V$ back into the Lagrangian, we find
\bea\label{LagrCan}
\mathrsfso{L}\sim \mathrm{Tr}\left(\left(U  D\bar{U} \right)^\dagger \left[\frac{1}{\frac{1}{2}\,\frac{1+s}{1-s}\,\mathrm{Id}+U\bar{U}\,{i\over 2}\,\mathcal{R}}\right]\, \left(U  D\bar{U} \right)\right)
\eea
An additional simplification occurs, if we use the formulation of the model where $U$ is an $n\times n$-matrix. In that case the unitary gauge implies that $U$ is unitary, so that, apart from $\bar{U} U=\mathds{1}_n$ we have $U \bar{U}=\mathds{1}_n$, and as a result one can forget the projector in the denominator in~(\ref{LagrCan}), arriving at the simple formula
\begin{empheq}[box=\fbox]{align}
\hspace{1em}\vspace{1em}
\label{LagrCan2}
\mathrsfso{L}= \mathrm{Tr}\left(\left(U  D\bar{U} \right)^\dagger \,r_s^{-1} \left(U  D\bar{U} \right)\right)\,,\quad\quad  U\in U(n)\,.\quad
\end{empheq}
Of course in this case the complications are absorbed in the enlarged gauge field $\mathcal{A}$.

\end{appendices}

\makeatletter
\renewcommand\@biblabel[1]{#1.}
\makeatother

\vspace{-0.2cm}
{ \setlength{\bibsep}{0.0pt}

  }

\end{document}